\documentclass[aps]{revtex4}
\usepackage{graphicx}
\usepackage{amsmath}
\usepackage{amssymb}
\begin{document}

\title{Color defects in a gauge condensate}
\author{Vladimir Dzhunushaliev
\footnote{Senior Associate of the Abdus Salam ICTP}} 
\email{dzhun@krsu.edu.kg} \affiliation{Dept. Phys. and Microel. 
Engineer., Kyrgyz-Russian Slavic University, Bishkek, Kievskaya Str. 
44, 720021, Kyrgyz Republic}


\begin{abstract}
The model of an approximate non-perturbative calculations in the SU(3) gauge theory is offered. This approach is based on the separation of initial degrees of freedom on ordered and disordered phases. The ordered phase is almost classical degrees of freedom, the disordered phase is completely quantum degrees of freedom. Using some approximations and simplifications for 2 and 4-points Green's functions an effective Lagrangian describing both phases from the SU(3) Lagrangian is obtained. The calculations show that ordered phase is squeezed by disordered phase into defects. These defects are: an infinite flux tube filled with longitudinal color electric and magnetic fields; a color electric hedgehog; a defect having either two color electric dipoles + two color magnetic dipoles or two color electric dipoles or two color magnetic dipoles. It assumed that the color defects are quantum excitations in a gauge condensate. The equations for the disordered phase are an analog of Ginzburg - Landau equation. 
\end{abstract}

\maketitle

\section{Introduction}

In QCD exists an open problem of the description of non-perturbative degrees of freedom. In the most known kind this problem is connected to the problem of quark confinement. In this work we offer an approximate model of non-perturbative quantization. We separate SU(3) degrees of freedom on two parts: the first one are almost classical degrees of freedom (ordered phase), the second one are entirely quantum degrees of freedom (disordered phase or gauge condensate). The ordered phase is described using classical field equations for a small gauge subgroup. A gauge condensate is presented by scalar fields with corresponding scalar field equations. In order to obtain the equations describing disordered phase we use several assumptions and simplifications for 2 and 4-points Green's functions. Roughly speaking we suppose that 2 and 4-points Green's functions can be simplified by the following way: $G_2 \approx \phi^2$ and $G_4 \approx G_2^2 + \alpha_1 G_1 + \alpha_2$ 
($\alpha_{1,2}$ are constants). 
\par 
In the modern language we derive field equations for the dimensions two and four condensates. As a consequence we have the possibility to describe the gauge condensate on the microscopical level as a field distribution. Using this idea we can investigate color defects in the gauge condensate. In our investigation the color defects are some defects which destroy the global spatial homogeneity of the disordered phase. It can be color flux tubes, color electric dipoles, color magnetic currents and so on. 
\par
The dimension two condensate $\left\langle A_\mu^2\right\rangle$ has received a great deal of attention in the last few years, see for example
\cite{Gubarev:2000eu}-\cite{Shakin:2004te}. The output of these investigations is that a non-vanishing condensate is favored as it lowers the vacuum energy. As a consequence of the existence of a non-vanishing condensate $\left\langle A_\mu^2\right\rangle$, a dynamical mass parameter for the gluons can be generated in the gauge fixed Lagrangian, 
see \cite{Verschelde:2001ia}, \cite{Dudal:2003by}.
\par
An interesting idea is implemented in the so-called ``cut-off'' model \cite{Engels:1989ph}, \cite{Rischke:1992uv} in which gluons having momenta smaller than a fixed value $K$ are bound inside non-perturbative structures. This idea is similar to the one we will use in this work: probably these gluons are in disordered phase which we investigate here. 
\par 
In the other context the condensate notion arises in the dual superconductor model \cite{DualSuperconductor} where the vacuum of a non--Abelian gauge model may be regarded as a medium of condensed Abelian monopoles. The monopole condensate squeezes the chromoelectric flux (coming from the quarks) into a flux tube due to the dual Meissner effect. In this work we show that the gauge condensate squeezes the color electric and magnetic fields into defects considered here.
\par 
In Ref. \cite{jia} it is shown that it is possible to arrive at an effective dual Abelian-Higgs model, the dual and relativistic version of Ginzburg-Landau model for superconductor, from $SU(2)$ Yang-Mills theory based on the Faddeev-Niemi connection decomposition and the order-disorder assumptions for the gauge field. 

\section{Heisenberg's non-perturbative quantization for QCD}

In this subsection we will apply a version of Heisenberg's non-perturbative quantization
method \cite{heisenberg} to QCD.  The classical SU(3) Yang-Mills equations are
\begin{equation}
    \partial_\nu \mathcal F^{B\mu\nu} = 0
\label{sec1-10}
\end{equation}
where $\mathcal F^B_{\mu \nu} = \partial_\mu \mathcal A^B_\nu -
\partial_\nu \mathcal A^B_\mu + g f^{BCD} \mathcal A^C_\mu \mathcal A^D_\nu$
is the field strength; $B,C,D = 1, \ldots ,8$ are the SU(3) color indices;
$g$ is the coupling constant; $f^{BCD}$ are the structure constants for
the SU(3) gauge group. In quantizing the system given in Eqs. (\ref{sec1-10}) - via Heisenberg's non-perturbative method one first replaces the classical fields by field operators $\mathcal A^B_{\mu} \rightarrow \widehat{\mathcal A}^B_\mu$. This yields the
following differential equations for the operators
\begin{equation}
    \partial_\nu \widehat {\mathcal F}^{B\mu\nu} = 0.
\label{sec1-20}
\end{equation}
These nonlinear equations for the field operators of the nonlinear quantum fields can be used to determine expectation values for the field operators
$\widehat {\mathcal A}^B_\mu$, where
$\langle \cdots \rangle = \langle Q | \cdots | Q \rangle$ and
$| Q \rangle$ is some quantum state. One can also use these
equations to determine the expectation values of operators
that are built up from the fundamental operators
$\widehat {\mathcal A}^B_\mu$. For example, the ``electric'' field
operator, $\widehat {\mathcal E}^B_z = \partial _0 \widehat {\mathcal A}^B_z -
\partial _z \widehat {\mathcal A}^B_0 + g f^{BCD} \mathcal A^C_0 \mathcal A^D_z$
giving the expectation $\langle \widehat {\mathcal E}^B_z \rangle$.
The simple gauge field expectation values,
$\langle \mathcal{A}_\mu (x) \rangle$, are obtained by average Eq. \eqref{sec1-20} over some quantum state $| Q \rangle$
\begin{equation}
  \left\langle Q \left|
  \partial_\nu \widehat {\mathcal F}^{B\mu\nu}
  \right| Q \right\rangle = 0.
\label{sec1-30}
\end{equation}
One problem in using these equations to obtain expectation values like 
$\langle \mathcal A^B_\mu \rangle$, is that these equations involve not only powers or derivatives of $\langle \mathcal A^B_\mu \rangle$ ({\it i.e.} terms like 
$\partial_\alpha \langle \mathcal A^B_\mu \rangle$ or $\partial_\alpha
\partial_\beta \langle \mathcal A^B_\mu \rangle$) 
but also contain terms like 
$\mathcal{G}^{BC}_{\mu\nu} = \langle \mathcal A^B_\mu \mathcal A^C_\nu \rangle$. 
Starting with Eq. \eqref{sec1-30} one can generate an operator differential equation for the product $\widehat {\mathcal A}^B_\mu \widehat {\mathcal A}^C_\nu$ thus allowing the determination of the Green's function $\mathcal{G}^{BC}_{\mu\nu}$
\begin{equation}
  \left\langle Q \left|
  \widehat {\mathcal A}^B(x) \partial_{y\nu} \widehat {\mathcal F}^{B\mu\nu}(x)
  \right| Q \right\rangle = 0.
\label{sec1-40}
\end{equation}
However this equation will in turn contain other, higher order Green's functions. Repeating these steps leads to an infinite set of equations connecting Green's functions of ever increasing order. This construction, leading to an infinite set of coupled, differential equations, does not have an exact, analytical solution
and so must be handled using some approximation.
\par
Operators which are involved in equation \eqref{sec1-20} are only well determined if there is a Hilbert space of quantum states. Thus we need to look into the question of the definition of the quantum states $| Q \rangle$ in the above construction. The resolution to this problem is as follows: There is an one-to-one correspondence between a given quantum state $| Q \rangle$ and the infinite set of quantum expectation values over any product of field operators,
$\mathcal{G}^{mn \cdots}_{\mu\nu \cdots}(x_1, x_2 \ldots) =
\langle Q | \mathcal A^m_\mu (x_1) \mathcal A^n_\nu (x_2) \ldots
| Q \rangle$. So if all the Green's functions
-- $\mathcal{G}^{mn \cdots}_{\mu\nu \cdots}(x_1, x_2 \ldots)$ --
are known then the quantum states, $| Q \rangle$ are known, \textit{i.e.} the action of 
$| Q \rangle$ on any product of field operators
$\widehat {\mathcal A}^m_\mu (x_1) \widehat {\mathcal A}^n_\nu (x_2) \ldots$
is known. The Green's functions are determined from the above,
infinite set of equations (following Heisenberg's idea).
\par
Another problem associated with products of field operators like 
$\widehat {\mathcal A}^m_\mu (x) \widehat {\mathcal A}^n_\nu (x)$
which occur in Eq. \eqref{sec1-20} is that the two operators occur at the same point. For \textit{non-interacting} field it is well known that such products have a singularity. In this paper we are considering \textit{interacting} fields so it is not known if a singularity would arise for such products of operators evaluated at the same point. Heisenberg in his investigations of a quantized nonlinear spinor field repeatedly underscored that in a quantum field theory with strong interaction the singularities of
propagators can be essentially smoothed out. Physically it is hypothesized that there are situations in interacting field theories where these singularities do not occur
({\it e.g.} for flux tubes in Abelian or non-Abelian theory
quantities like the ``electric'' field inside the tube, $\langle
\mathcal E^a_z \rangle < \infty$, and energy density $\varepsilon
(x) = \langle (\mathcal E^a_z)^2 \rangle < \infty$ are
nonsingular). Here we take as an assumption that such
singularities do not occur.

\section{Infinite color flux tube defect.}
\label{fluxtube}

In this section we will present a cylindrically symmetric color defect in a gauge condensate -- infinite flux tube between dyon and antidyon. We reiterate that we have not found an exact analytical method for solving the full equations for all the Green's functions. Our basic approach in this case is to give some physically reasonable scheme for cutting off the infinite set of equations for the Green's functions. In addition our approximate calculations will involve the decomposition of the initial gauge group into a smaller gauge group: 
$SU(3) \rightarrow SU(2) \times U(1)$. Physically we will find that this reduction splits the initial degrees of freedom (the $SU(3)$ gauge potential components) into $SU(2) \times U(1)$ and coset components. The $SU(2) \times U(1)$ components will represent the ordered phase while the coset components, which will be represented via an effective scalar field, will represent the disordered phase. After these approximations one can perform analytical calculations which suggest the formation of a color electric flux tube. This is of interest since an important feature of the confinement phenomenon is the formation of electric flux tubes between the confined quarks.
\par 
We suppose that quantum SU(3) gauge fields in quantum chromodynamics can be decomposed on a gauge condensate (ordered phase) and non-perturbative fluctuations (disordered phase). We suppose as well as that the flux tube is an axially symmetric defect in a gauge condensate. 

\subsection{Reduction to a small subgroup}

First we define the reduction of $SU(3)$ to $SU(2) \times U(1)$ following the conventions of Ref. \cite{kondo}. Starting with the SU(3) gauge group with generators $T^B$, we define the $SU(3)$ gauge fields, $\mathcal{A}_\mu=\mathcal{A}^B_\mu T^B$. $SU(2) \times U(1)$ is a subgroup of $SU(3)$ and $SU(3)/(SU(2) \times U(1))$ is a coset. Then the gauge field $\mathcal{A}_\mu$ can be decomposed as
\begin{eqnarray}
  \mathcal{A}_\mu & = & \mathcal{A}^B_\mu T^B = A^a_\mu T^a + A^8_\mu T^8 + A^m_\mu T^m ,
\label{2a-10}\\
  A^a_\mu & \in & SU(2), \quad A^8_\mu \in U(1) \quad 
  \text{and} \quad A^m_\mu \in SU(3)/\left( SU(2) \times U(1) \right)
\label{2a-20}
\end{eqnarray}
where the indices $a,b,c \ldots = 1,2,3$ belongs to the subgroup SU(2) and 
$m,n, \ldots =4,5,6,7$ to the coset $SU(3)/\left( SU(2) \times U(1) \right)$; $B$ are SU(3) indices. Based on this the field strength can be decomposed as
\begin{equation}
  \mathcal{F}^B_{\mu\nu} T^B = \mathcal{F}^a_{\mu\nu}T^a + \mathcal{F}^8_{\mu\nu}T^8 +
  \mathcal{F}^m_{\mu\nu}T^m
\label{2a-30}
\end{equation}
where
\begin{align}
  \mathcal{F}^a_{\mu\nu} & = F^a_{\mu\nu} + \Phi^a_{\mu\nu}
  & \in &SU(2) ,
\label{2a-40}\\
  F^a_{\mu\nu} & = \partial_\mu A^a_\nu - \partial_\nu A^a_\mu +
  g f^{abc} A^b_\mu A^c_\nu & \in &SU(2) ,
\label{2a-50}\\
  \Phi^a_{\mu\nu} & = g f^{amn} A^m_\mu A^n_\nu & \in &SU(2) ,
\label{2a-60}\\
  \mathcal{F}^8_{\mu\nu} & = h_{\mu\nu} + g f^{8mn} A^m_\mu A^n_\nu 
  & \in &U(1) ,
\label{2a-70}\\
  h_{\mu\nu} & = \partial_\mu b_\nu - \partial_\nu b_\mu
  & \in &U(1) , \quad b_\mu = A^8_\mu, 
\label{2a-80}\\  
	\mathcal{F}^m_{\mu\nu} & = F^m_{\mu\nu} + G^m_{\mu\nu} &
  \in &SU(3)/\left( SU(2) \times U(1) \right) , 
\label{2a-90}\\
  F^m_{\mu\nu} & = \partial_\mu A^m_\nu - \partial_\nu A^m_\mu 
  & \in &SU(3)/\left(SU(2) \times U(1) \right) ,
\label{2a-100}\\
  G^m_{\mu\nu} & = g f^{mnb}
  \left(
  A^n_\mu A^b_\nu - A^n_\nu A^b_\mu
  \right) + g f^{mn8} 
  \left(
  A^n_\mu b_\nu - A^n_\nu b_\mu
  \right)  
  & \in &SU(3)/\left( SU(2) \times U(1) \right)
\label{2a-110}
\end{align}
where $f^{abc}$ are the structure constants of SU(2). The SU(3) Yang-Mills field equations can be decomposed as
\begin{eqnarray}
  d_\nu \left( F^{a\mu\nu} +\Phi^{a\mu\nu} \right) & = &
  -g f^{amn} A^m_\nu
  \left(
  F^{n\mu\nu} + G^{n\mu\nu}
  \right) ,
\label{2a-120}\\
  D_\nu \left( F^{m\mu\nu} + G^{m\mu\nu} \right) & = &
  - g f^{mnb}
  \left[
  A^n_\nu \left( h ^{b\mu\nu} + \Phi^{b\mu\nu} \right) - 
  a^b_\nu \left( F^{n\mu\nu} + G^{n\mu\nu} \right)
  \right] - 
\nonumber \\
  &&
  g f^{mn8} \left[
  	A^n_\nu h^{\mu \nu} - b_\nu \left( F^{n\mu\nu} + G^{n\mu\nu} \right)
  \right],
\label{2a-130}\\
	\partial_\nu \left(
		h^{\mu \nu} + g f^{8mn} A^{m \mu} A^{n \nu}
	\right) &=& -g^{8mn} A^m_\nu 
  \left(
  F^{n\mu\nu} + G^{n\mu\nu}
  \right)
\label{2a-132}
\end{eqnarray}
where $d_\nu [\cdots]^a = \partial_\nu [\cdots]^a + f^{abc} a^b_\nu [\cdots]^c$ is the covariant derivative on the subgroup SU(2) and
$D_\nu [\cdots]^m = \partial_\nu [\cdots]^m + f^{mnA} A^n_\nu [\cdots]^A$. The Heisenberg non-perturbative quantization procedure gives us 
\begin{eqnarray}
  d_\nu \left( \widehat h^{a\mu\nu} + \widehat \Phi^{a\mu\nu} \right) & = &
  -g f^{amn} \widehat A^m_\nu
  \left(
  \widehat F^{n\mu\nu} + \widehat G^{n\mu\nu}
  \right) ,
\label{2a-140}\\
  D_\nu \left( \widehat F^{m\mu\nu} + \widehat G^{m\mu\nu} \right) & = &
  - g f^{mnb}
  \left[
  \widehat A^n_\nu \left( \widehat h ^{b\mu\nu} + \widehat \Phi^{b\mu\nu} \right) -
  \widehat a^b_\nu \left( \widehat F^{n\mu\nu} + \widehat G^{n\mu\nu} \right)
  \right] - 
\nonumber \\
  &&
  g f^{mn8} \left[
  	\widehat A^n_\nu \widehat h^{\mu \nu} - 
  	\widehat b_\nu \left( \widehat F^{n\mu\nu} + \widehat G^{n\mu\nu} \right)
  \right], 
\label{2a-150}\\
	\partial_\nu \left(
		\widehat h^{\mu \nu} + g f^{8mn} \widehat A^{m \mu} \widehat A^{n \nu}
	\right) &=& -g^{8mn} \widehat A^m_\nu 
  \left(
  \widehat F^{n\mu\nu} + \widehat G^{n\mu\nu}
  \right)
\label{2a-160}
\end{eqnarray}

\subsection{Basic assumptions}
\label{basic_assumptions}

It is evident that an exact quantization is impossible for Eq's. \eqref{2a-140}-\eqref{2a-160}. Thus we look for some simplification in order to obtain equations which can be analyzed. Our basic aim is to cut off the set of infinite equations using some simplifying assumptions. For this purpose we will propose simple but physically reasonable ans\"atzen for the 2 and 4-points Green's functions -- 
$\left\langle A^m_\mu(y) A^n_\nu(x)\right\rangle$, 
$\left\langle A^a_\alpha (x) A^b_\beta (y)A^m_\mu(z) A^n_\nu(u) \right\rangle$, 
$\left\langle A^8_\alpha (x) A^8_\beta (y)A^m_\mu(z) A^n_\nu(u) \right\rangle$
and 
$\left\langle A^m_\alpha (x) A^n_\beta (y) A^p_\mu(z) A^q_\nu(u) \right\rangle$ 
-- which are involved in averaged SU(3) Lagrangian. As was mentioned earlier we assume that there are two phases
\begin{enumerate}
\item The gauge field components $A^{a,8}_\mu$ ($A^a_\mu \in SU(2), A^8_\mu \in 					U(1)$) belonging to the small subgroup $SU(2) \times U(1)$ are in an ordered phase. Mathematically this means that
\begin{equation}
  \left\langle a^a_\mu (x) \right\rangle  =(a^a _{\mu} (x))_{cl}.
\label{2b-10}
\end{equation}
      The subscript means that this is a classical field. Thus we are
      treating these components as effectively classical gauge fields
      in the first approximation.
\item 
\label{2}
The gauge field components $A^m_\mu$ ($m=4,5, ... , 7$ and
$A^m_\mu \in SU(3)/\left( SU(2) \times U(1) \right)$) belonging to the coset 
$SU(3)/\left( SU(2) \times U(1) \right)$ are in a disordered phase ({\it i.e.} they form a gauge condensate), but have non-zero energy. In mathematical terms this means that
\begin{equation}
  \left\langle A^m_\mu (x) \right\rangle = 0.
\label{2b-20}
\end{equation}
According to lattice calculations the main contribution at calculations of any quantum quantities in QCD is brought with field configurations having Abelian monopoles. We use this fact to say that the contribution from the $A_0^m$ component of gauge potential is negligible in comparing with the contribution from the $A^m_\mu$ components $\mu = 1,2,3$ 
\begin{equation}
  \left\langle A^m_0 (x) A^n_0 (y) \right\rangle \ll 
  \left\langle A^m_\mu (x) A^n_\nu (y) \right\rangle
\label{2b-25}
\end{equation}
where $\mu, \nu = 1,2,3$ are the spatial indices. After this remark we define an approximate expression for the 2-point Green's function as follows 
\begin{eqnarray}
  \left\langle A^m_\mu (x) A^n_\nu (y) \right\rangle &=& - \eta_{\mu \nu} 
  f^{mpa} f^{npb} \phi^a(x) \phi^b(y),
\label{2b-30}\\
	\left\langle A^m_0 (x) A^n_0 (y) \right\rangle & \approx & 0
\label{2b-31}
\end{eqnarray}
here $\phi^a$ is a real SU(2) triplet scalar fields. Thus we have replaced the coset gauge fields in 2-points Green's function by effective scalar fields, which will be the scalar field in our effective $SU(2) \times U(1)$-scalar system. 
\item 
We suppose that the Green's functions with odd numbers of $A^m_\mu$ are zero 
\begin{equation}
  \left\langle 
  	\overbrace{A^m_\mu (x) \ldots A^n_\nu (y)}^{\text{odd number}} 
  \right\rangle = 0 
\label{2b-32}
\end{equation}
\item The correlations between the ordered (classical) and disordered (quantum) phases have the following forms 
\begin{eqnarray}
  \left\langle A^m_\mu(x) A^n_\nu(x) b_\alpha (z) \right\rangle &=&
  k_1 b_\alpha (z) \left\langle A^m_\mu(x) A^n_\nu(x) \right\rangle ,
\label{2b-40}\\
  \left\langle A^m_\mu(x) A^n_\nu(y) b_\alpha (z) b_\beta (u) \right\rangle &=&
  k_1^2 b_\alpha (z) b_\beta (u) \left\langle A^m_\mu(x) A^n_\nu(x) \right\rangle + 
  \delta^{mn} M_{1, \mu \nu} b_\alpha (z) b_\beta (u) ,
\label{2b-50}\\
  \left\langle A^m_\mu(x) A^n_\nu(y) A^a_\alpha (z) A^b_\beta (u) \right\rangle &=&
  k_2^2 A^a_\alpha (z) A^b_\beta (u) \left\langle A^m_\mu(x) A^n_\nu(x) \right\rangle + 
  \delta^{mn} M_{2, \mu \nu} A^a_\alpha (z) A^b_\beta (u) ,
\label{2b-60}
\end{eqnarray}
where $k_{1,2}$ are constants; $M_{1,2, \mu \nu}$ are matrices which will destroy the gauge invariance of a final Lagrangian. 
\item The 4-points Green's functions can be decomposed by the following way. For $A^m_\mu$ gauge potentials 
\begin{eqnarray}
  \left\langle A^m_\mu(x) A^n_\nu(y) A^p_\alpha(z) A^q_\beta(u) \right\rangle &=&
  \lambda_1 \biggl[ \left\langle A^m_\mu(x) A^n_\nu(y) \right\rangle 
	  \left\langle A^p_\alpha(z) A^q_\beta(u) \right\rangle + 
  \biggl.,
\nonumber \\
	  &&
	  \frac{\mu_1^2}{4} \left(
	  	\delta^{mn} \eta_{\mu \nu} \left\langle A^p_\alpha(z) A^q_\beta(u) \right\rangle + 
	  	\delta^{pq} \eta_{\alpha \beta} \left\langle A^m_\mu(x) A^n_\nu(y) \right\rangle
	  \right) + 
\nonumber \\
	  &&
	  \frac{\mu_1^2}{4} \left(
	  	\delta^{mp} \eta_{\mu \alpha} \left\langle A^n_\nu(y) A^q_\beta(u) \right\rangle + 
	  	\delta^{nq} \eta_{\nu \beta} \left\langle A^m_\mu(x) A^p_\alpha(y) \right\rangle
	  \right) + 	
\nonumber \\
	  &&
	  \frac{\mu_1^2}{4} \left(
	  	\delta^{mq} \eta_{\mu \beta} \left\langle A^n_\nu(y) A^p_\alpha(u) \right\rangle + 
	  	\delta^{np} \eta_{\nu \alpha} \left\langle A^m_\mu(x) A^q_\beta(y) \right\rangle
	  \right) +    ,
\nonumber \\
	  &&
	 \biggl.
	  \frac{\mu_1^4}{16} \left( 
	  \delta^{mn} \eta_{\mu \nu} \delta^{pq} \eta_{\alpha \beta} + 
	  \delta^{mp} \eta_{\mu \alpha} \delta^{nq} \eta_{\nu \beta} +
	  \delta^{mq} \eta_{\mu \beta} \delta^{np} \eta_{\nu \alpha} 
	  \right)
	  \biggl] 
\label{2b-70}
\end{eqnarray}
where $\lambda_1, \mu_1$ are constants. 
\end{enumerate}

\subsection{Derivation of an effective Lagrangian}

For our quantization procedure we will take the expectation of the Lagrangian rather than for the equations of motions. Thus we will obtain an effective Lagrangian rather than approximate equations of motion. The Lagrangian we obtain from the original SU(3) pure gauge theory will turn out to be an effective $SU(2) \times U(1)$ Yang-Mills-Higgs system. The averaging SU(3) Lagrangian is
\begin{equation}
  \mathcal{L}_{eff} = - \frac{1}{4}\left\langle 												              				\mathcal{F}^A_{\tilde{\mu}\tilde{\nu}}\mathcal{F}^{A\tilde{\mu}\tilde{\nu}} 
  \right\rangle =   - \frac{1}{2}\left\langle 												              					\mathcal{F}^A_{0\nu}\mathcal{F}^{A0\nu} \right\rangle 
  - \frac{1}{4}\left\langle \mathcal{F}^A_{\mu\nu}\mathcal{F}^{A\mu\nu}\right\rangle 
\label{2c-10}
\end{equation}
where $\tilde{\mu}, \tilde{\nu} = 0,1,2,3$. According to item \ref{2} of section \ref{basic_assumptions}
\begin{equation}
  \left\langle \mathcal{F}^A_{0\nu}\mathcal{F}^{A0\nu} \right\rangle \ll 
  \left\langle \mathcal{F}^A_{\mu\nu}\mathcal{F}^{A\mu\nu}\right\rangle 
\label{2c-15}
\end{equation}
since $\mathcal{F}^A_{0\nu} = - \partial_\mu A_0 + g f^{ABC} A^B_0 A^C_\mu$ 
(we consider the stationary fields only), 
$\mathcal{F}^A_{\mu\nu} = \partial_\mu A_\nu - \partial_\mu A_\nu + g f^{ABC} A^B_\mu A^C_\nu$ and correspondingly 
$\left\langle \mathcal{F}^A_{0\nu}\mathcal{F}^{A0\nu} \right\rangle$ has terms like 
$\left\langle \mathcal{A}^B_{0} \mathcal{A}^{B0} \right\rangle$ but 
$\left\langle \mathcal{F}^A_{\mu\nu}\mathcal{F}^{A0\mu\nu} \right\rangle$ -- 
$\left\langle \mathcal{A}^B_{\mu} \mathcal{A}^{B\mu} \right\rangle$. It means that the magnetic fields generating by the magnetic monopoles are more important by quantization in comparing with the electric fields. Therefore
\begin{equation}
  \mathcal{L}_{eff} \approx 
  - \frac{1}{4}\left\langle \mathcal{F}^A_{\mu\nu}\mathcal{F}^{A\mu\nu}\right\rangle = 
	- \frac{1}{4} \left( 
	\left\langle \mathcal{F}^a_{\mu\nu}	\mathcal{F}^{a\mu\nu} \right\rangle + 
	\left\langle \mathcal{F}^8_{\mu\nu}	\mathcal{F}^{8\mu\nu} \right\rangle + 
	\left\langle \mathcal{F}^m_{\mu\nu} \mathcal{F}^{m\mu\nu} \right\rangle \right)
\label{2c-16}
\end{equation}
In Eq. \eqref{2c-16}  $\mathcal{F}^{a\mu\nu}, \mathcal{F}^{8\mu\nu}$ and $\mathcal{F}^{m\mu\nu}$ are defined by equations \eqref{2a-40}-\eqref{2a-110}.

\subsubsection{Calculation of
$\left\langle \mathcal{F}^a_{\mu\nu} \mathcal{F}^{a\mu\nu} \right\rangle$}
\label{1a}

We begin by calculating the first term on the r.h.s. of equation
\eqref{2c-16}
\begin{equation}
  \left\langle \mathcal{F}^a_{\mu\nu} \mathcal{F}^{a\mu\nu} \right\rangle =
  \left\langle F^a_{\mu\nu} F^{a\mu\nu} \right\rangle +
  2\left\langle F^a_{\mu\nu} \Phi^{a\mu\nu} \right\rangle +
  \left\langle \Phi^a_{\mu\nu} \Phi^{a\mu\nu} \right\rangle .
\label{2c1-10}
\end{equation}
Immediately we see that the first term on the r.h.s. of equation \eqref{2c1-10} is the SU(2) Lagrangian as we assume that $A^a_\mu$ and $F^a_{\mu\nu}$ are effectively classical quantities and consequently
\begin{equation}
  \left\langle F^a_{\mu\nu} F^{a\mu\nu} \right\rangle =
  F^a_{\mu\nu} F^{a\mu\nu} .
\label{2c1-20}
\end{equation}
The second term in equation \eqref{2c1-10} vanishes as $F^a_{\mu \nu}$ is the antisymmetrical tensor but $\left\langle \Phi^a_{\mu \nu} \right\rangle$ is the symmetrical one (see Eq. \eqref{2b-30}). 
\par
The last term which is quartic in the coset gauge fields can be calculated using 
\eqref{2b-70}
\begin{equation}
  f^{amn} f^{apq} \left\langle A^m_\mu A^n_\nu A^{p\mu} A^{q \nu} \right\rangle =
  \frac{9}{8} \lambda_1 g^2 \left(
  	\phi^a \phi^a - \mu_1^2
  \right)^2.
\label{2c1-40}
\end{equation}
Up to this point the SU(2) part of the Lagrangian is
\begin{equation}
  \left\langle \mathcal{F}^a_{\mu\nu} \mathcal{F}^{a\mu\nu} \right\rangle =
  \left( F^a_{\mu\nu} F^{a\mu\nu} \right) +
  \frac{9}{8} \lambda_1 g^2 \left(
  	\phi^a \phi^a - \mu_1^2
  \right)^2.
\label{2c1-50}
\end{equation}

\subsubsection{Calculation of
$\left\langle \mathcal{F}^m_{\mu\nu} \mathcal{F}^{m\mu\nu} \right\rangle$}
\label{1b}

Next we work on the coset part of the Lagrangian
\begin{equation}
\begin{split}
  \left\langle \mathcal{F}^m_{\mu\nu} \mathcal{F}^{m\mu\nu} \right\rangle = &
  \left\langle
		F^m_{\mu\nu} F^{m\mu\nu}
  \right\rangle + 
  2 g f^{mna} \left\langle 
  	F^m_{\mu\nu} \left(
  		A^{a \nu} A^{n \mu} - A^{a \mu} A^{n \nu}
  	\right)
  \right\rangle + 
  \\
  &
    2 g f^{mn8} \left\langle 
  	F^m_{\mu\nu} \left(
  		b^\nu A^{n \mu} - b^\mu A^{n \nu}
  	\right)
  \right\rangle + 
  g^2 f^{mna} f^{mpb} \left\langle 
  	\left( A^a_\nu A^n_\mu - A^a_\mu A^n_\nu \right)
  	\left( A^{b\nu} A^{p \mu} - A^{b \mu} A^{p \nu} \right)
  \right\rangle + 
  \\
  &
  g^2 f^{mn8} f^{mp8} \left\langle 
  	\left( b_\nu A^n_\mu - b_\mu A^n_\nu \right)
  	\left( b^\nu A^{p \mu} - b^\mu A^{p \nu} \right)
  \right\rangle + 
  \\
  &
  g^2 f^{mna} f^{mp8} \left\langle 
  \left( A^a_\nu A^n_\mu - A^a_\mu A^n_\nu \right)
  \left( b^\nu A^{p \mu} - b^\mu A^{p \nu} \right)
  \right\rangle .
\end{split}
\label{2c2-10}
\end{equation}
After the calculations we have 
\begin{eqnarray}
	\left\langle
		F^m_{\mu\nu} F^{m\mu\nu}
  \right\rangle &=& 
  -4 \left( \partial_\mu \phi^a \right) \left( \partial^\mu \phi^a \right) ,
\label{2c2-11}\\
	g^2 f^{mna} f^{mpb} 
	\left\langle 
		\left( A^a_\nu A^n_\mu - A^a_\mu A^n_\nu \right)
		\left( A^{b \nu} A^{p \mu} - A^{b \mu} A^{p \nu} \right)
	\right\rangle
	&=& 
  - g^2 k_2^2 \left( \phi^b \phi^b \right) 
  A^a_\mu A^{a \mu} + 2 \left( M^2 \right)^{\mu \nu} A^a_\mu A^a_\nu ,
\label{2c2-12}\\
	g^2 f^{mn8} f^{mp8} 
	\left\langle 
		\left( b_\nu A^n_\mu - b_\mu A^n_\nu \right)
		\left( b^\nu A^{p \mu} - b^\mu A^{p \nu} \right)
	\right\rangle
	&=& 
  - 3 g^2 k_1^2 \left( \phi^a \phi^a \right) 
  b_\mu b^\mu + 2 \left( m^2 \right)^{\mu \nu} b_\mu b_\nu ,
\label{2c2-13}\\
	2 g f^{mna} 
	\left\langle 
		F^m_{\mu \nu}
		\left( A^{a \nu} A^{n \mu} - A^{a \mu} A^{n \nu} \right)
	\right\rangle
	&=& 
  - 2 g k_2 f^{abc} \left( \partial_\mu \phi^a \right) A^{b \mu} \phi^c ,
\label{2c2-14}\\
	g^2 f^{mna} f^{mp8} 
	\left\langle 
		\left( A^a_\nu A^n_\mu - A^a_\mu A^n_\nu \right)
		\left( b^\nu A^{p \mu} - b^\mu A^{p \nu} \right)
	\right\rangle
	&=& 0 ,
\label{2c2-15}\\
	2g f^{mn8} \left\langle 
		F^m_{\mu \nu} \left( b^\nu A^{n \mu} - b^\mu A^{n \nu} \right)
	\right\rangle &=& 0 
\label{2c2-17}
\end{eqnarray}
where 
\begin{eqnarray}
	\left( M^2 \right)^{\mu \nu} &=& g^2 \left(
		M^\alpha_{2, \alpha} \eta^{\mu \nu} - M^{\mu \nu}_2
	\right),
\label{2c2-15a}\\
	\left( m^2 \right)^{\mu \nu} &=& 3 g^2 \left(
		M^\alpha_{1, \alpha} \eta^{\mu \nu} - M^{\mu \nu}_1
	\right).
\label{2c2-15b}
\end{eqnarray}
In what follows we assign these matrices as diagonal 
\begin{eqnarray}
	\left( M^2 \right)^{\mu \nu} &=& M^2_\mu \eta^{\mu \nu},
\label{2c2-15c}\\
	\left( m^2 \right)^{\mu \nu} &=& m^2_\mu \eta^{\mu \nu}.
\label{2c2-15d}
\end{eqnarray}
on $\mu$ there is no summing, $M^2_\mu$ and $m^2_\mu$ are constants. 
Finally 
\begin{equation}
\begin{split}
   \left\langle
		\mathcal F^m_{\mu\nu} \mathcal F^{m\mu\nu}
  \right\rangle = &
  -4 \left( \partial_\mu \phi^a \right) \left( \partial^\mu \phi^a \right) 
  - 2 g k_2 f^{abc} \left( \partial_\mu \phi^a \right) A^{b \mu} \phi^c 
  - g^2 k_2^2 \left( \phi^b \phi^b \right) 
  A^a_\mu A^{a \mu} 
  - 3 g^2 k_1^2 \left( \phi^a \phi^a \right) 
  b_\mu b^\mu + 
  \\
  &
  2 \left( m^2 \right)^{\mu \nu} b_\mu b_\nu + 
  2 \left( M^2 \right)^{\mu \nu} A^a_\mu A^a_\nu .
\end{split}
\label{2c2-16}
\end{equation}

\subsubsection{Calculation of
$\left\langle \mathcal{F}^8_{\mu\nu} \mathcal{F}^{8 \mu\nu} \right\rangle$}
\label{1c}

We end by calculating the second term on the r.h.s. of equation
\eqref{2c-16}
\begin{equation}
  \left\langle \mathcal{F}^8_{\mu\nu} \mathcal{F}^{8 \mu\nu} \right\rangle =
  \left\langle h_{\mu\nu} h^{\mu\nu} \right\rangle +
  2 g \left\langle h_{\mu\nu} A^{m \mu} A^{n \nu} \right\rangle +
  g^2 f^{8mn} f^{8pq} 
  \left\langle 
  	A^m_\mu A^n_\nu A^{p \mu} A^{q \nu} 
  \right\rangle .
\label{2c3-10}
\end{equation}
The first term on the r.h.s. of \eqref{2c3-10} has the classical field $h_{\mu \nu}$ consequently 
\begin{equation}
  \left\langle h_{\mu\nu} h^{\mu\nu} \right\rangle = 
  h_{\mu\nu} h^{\mu\nu}.
\label{2c3-20}
\end{equation}
The second term in equation \eqref{2c3-10} vanishes as $h_{\mu \nu}$ is the antisymmetrical tensor but $\left\langle A^{m \mu} A^{n \nu} \right\rangle$ is the symmetrical one (see Eq. \eqref{2b-30}). The last term which is quartic in the coset gauge fields can be calculated using \eqref{2b-70}
\begin{equation}
  g^2 f^{8mn} f^{8pq} \left\langle A^m_\mu A^n_\nu A^{p\mu} A^{q \nu} \right\rangle =
  \frac{9}{8} \lambda_1 g^2 \left(
  	\phi^a \phi^a - \mu_1^2
  \right)^2.
\label{2c3-30}
\end{equation}
Thus 
\begin{equation}
  \left\langle \mathcal{F}^8_{\mu\nu} \mathcal{F}^{8\mu\nu} \right\rangle =
  h_{\mu\nu} h^{\mu\nu} +
  \frac{9}{8} \lambda_1 g^2 \left(
  	\phi^a \phi^a - \mu_1^2
  \right)^2.
\label{2c3-40}
\end{equation}

\subsection{An effective Lagrangian}

Thus after the quantization $A^m_\mu$ degrees of freedom we have 
\begin{equation}
\begin{split}
   \left\langle \mathcal{F}^A_{\mu\nu} \mathcal{F}^{A\mu\nu} \right\rangle = &
  F^a_{\mu\nu} F^{a\mu\nu} + h_{\mu\nu} h^{\mu\nu} - 
  4 \left(
  \partial_\mu \phi^a \partial^\mu \phi^a + 
  \frac{k_2}{2} g \epsilon^{abc} \partial_\mu \phi^a A^{b \mu} \phi^c + 
  \frac{1}{4} g^2 k_2^2 A^b_\mu A^{b \mu} \phi^c \phi^c
  \right) + 
  \\
  &
  \frac{9}{4} \lambda_1 g^2 \left( \phi^a \phi^a - \mu_1^2 \right)^2 
   - 3 k_1^2 g^2 \left( \phi^a \phi^a \right) b_\mu b^\mu + 
   2 \left( M^2 \right)^{\mu \nu} A^a_\mu A^a_\nu + 
   2 \left( m^2 \right)^{\mu \nu} b_\mu b_\nu 
\end{split}
\label{2c2-20}
\end{equation}
After the redefinitions $\phi^a \rightarrow \frac{\phi^a}{\sqrt{2}}$, 
$\mu_1 \rightarrow \frac{\mu_1}{\sqrt{2}}$, 
$\frac{9 g^2 \lambda_1}{4} \rightarrow \lambda_1$ we have the full averaged Lagrangian 
\begin{equation}
\begin{split}
	\mathcal L_{eff} = 
  - \frac{1}{4} \left\langle \mathcal{F}^A_{\mu\nu} \mathcal{F}^{A\mu\nu} \right\rangle = &
  - \frac{1}{4} F^a_{\mu\nu} F^{a\mu\nu} - \frac{1}{4} h_{\mu\nu} h^{\mu\nu} + 
  \frac{1}{2} \left( \tilde{D}_\mu \phi^a \right) 
  \left( \tilde{D}^\mu \phi^a \right) - 
  \frac{1}{8} g^2 k_2^2 A^b_\mu A^c_\mu \phi^c \phi^b - 
  \\
  &
  \frac{\lambda_1}{4} \left( \phi^a \phi^a - \mu_1^2 \right)^2 
   + \frac{3}{8}  k_1^2 g^2 \left( \phi^a \phi^a \right) b_\mu b^\mu - 
   \frac{1}{2} \left( M^2 \right)^{\mu \nu} A^a_\mu A^a_\nu - 
   \frac{1}{2} \left( m^2 \right)^{\mu \nu} b_\mu b_\nu 
\end{split}
\label{2c2-50}
\end{equation}
where 
$\tilde{D}_\mu \phi^a = \partial_\mu \phi^a + 
\frac{g}{2} k_2 \epsilon^{abc} A^{b \mu} \phi^c$. The original pure SU(3) gauge theory has been transformed into an $SU(2) \times U(1)$ gauge theory with broken gauge symmetry and coupled to an effective triplet scalar field. It is necessary to note that the coupling constant for an effective triplet scalar field is $\frac{g}{2} k_2$ not $g$. This Lagrangian we will investigate numerically on the next sections.
\par
Our final result given in \eqref{2c2-50} depends on several crucial assumptions.
In the next section we will investigate a non-topological flux tube solution to this
effective Lagrangian.

\subsection{Initial equations}

In this section we will investigate a \emph{non-topological} flux tube solution to the
effective Lagrangian \eqref{2c2-50}, \emph{i.e.} a cylindrically symmetric defect in a gauge condensate. We will see that the corresponding solution is an infinite flux tube filled with longitudinal and transversal electric and magnetic fields. It allows us to say that the obtained solution is the flux tube stretched between two color charge with chromoelectric and chromomagnetic fields. The separation between charge is infinite one. 
\par
We will start from the $SU(2) \times U(1)$ Yang - Mills - Higgs field equations of the Lagrangian \eqref{2c2-50} with broken gauge symmetry
\begin{eqnarray}
  D_\nu F^{a\mu\nu} &=& g \epsilon^{abc} \phi^b
  \tilde{D}^\mu \phi^c - \left( M^2 \right)^{\mu \nu} A^a_\nu - 
  \frac{1}{4} k_2^2 g^2 A^{b \mu} \phi^b \phi^a,
\label{2d-5}\\
	\partial_\nu h^{\mu \nu} &=& \frac{3}{4} k_1^2 g^2 
	\left( \phi^a \phi^a \right) b^\mu - 
	\left( m^2 \right)^{\mu \nu} b_\nu , 
\label{2d-10}\\
  \tilde{D}_\mu \tilde{D}^\mu \phi^a &=& -\lambda \phi^a
  \left(
  \phi^b \phi^b - \mu^2_1
  \right) + \frac{3}{4} g^2 k_1^2 b_\mu b^\mu \phi^a - 
  \frac{1}{4} g^2 k_2^2 A^b_\mu A^{a \mu} \phi^b
\label{2d-15}
\end{eqnarray}
here 
$F_{a\mu\nu} = \partial_\mu A^a_\nu - \partial_\nu A^a_\mu + 
g f^{abc} A^b_\mu A^c_\nu$ is the field tensor for the SU(2) gauge potential $A^a_\mu$; $a,b,c = 1,2,3$ are the color indices; 
$D_\nu [\cdots ]^a = \partial_\nu [\cdots ]^a + g f^{abc} A^b_\mu  [\cdots ]^c$ is the gauge derivative; $\phi^a$ is the Higgs field; $\lambda_1 , g$ and $\mu_1$ are some constants; $\left( M^2 \right)^{\mu \nu}$ and $\left( m^2 \right)^{\mu \nu}$ are masses matrices which destroys the gauge invariance of the Yang - Mills - Higgs theory. In the present case we are postulating that the masses of the two gauge bosons are different. This was done in order to find electric flux tube solutions for the system \eqref{2d-5}-\eqref{2d-15}. For equal masses electric flux tubes did not exist for the ans\"atz used. In numerically solving \eqref{2d-5}-\eqref{2d-15} the two mass come out close to one another, but are not equal. 
\par
We will use the following ans\"atz
\begin{equation}
\begin{split}
    &A^1_t(\rho) = \frac{f(\rho)}{g} ; \quad A^2_z(\rho) = \frac{v(\rho)}{g} ;
    \quad \phi^3(\rho) = \frac{\phi(\rho)}{g}; 
\\
		&
    b_\mu = \frac{1}{g} \left\{ h(\rho), 0, 0, w(\rho) \right\}; 
    \quad 
    \left( M^2 \right)^{\mu \nu} =\mathrm{diag} \left\{ M_0, M_1, 0, 0 \right\}; 
    \quad 
    \left( m^2 \right)^{\mu \nu} =
    \mathrm{diag} \left\{ m_0, 0, 0, \frac{m_3}{\rho^2} \right\}  
\label{2d-20}
\end{split}
\end{equation}
here $z, \rho , \varphi$ are cylindrical coordinates. The color electric and magnetic fields are 
\begin{eqnarray}
  E^3_z &=& \frac{1}{g} fv, \quad E^1_\rho = - \frac{1}{g} f', \quad 
  H^2_\varphi = \frac{1}{g} v' ,
\label{2d-24}\\
  E^8_\rho &=& \frac{1}{g} h' , \quad H^8_z = \frac{1}{g} \frac{w'}{\rho}.
\label{2d-28}
\end{eqnarray}
Substituting this into the Yang - Mills - Higgs equations \eqref{2d-5}-\eqref{2d-15} gives us
\begin{eqnarray}
    v'' + \frac{v'}{x} &=& v \left( \phi^2 - f^2 - M^2_1 \right),
\label{2d-30}\\
    f'' + \frac{f'}{x} &=& f \left( \phi^2 + v^2 - M^2_0 \right),
\label{2d-40}\\
    \phi'' + \frac{\phi'}{x} &=& \phi \left[ - f^2 + v^2
    	+ \lambda_1 \left( \phi^2 - \mu^2_1 \right) + 
    	\frac{3}{4} k_1^2 \left( 
    	- h^2 + \frac{w^2}{\rho^2} \right)
    \right]
\label{2d-50}\\
    h'' + \frac{h'}{x} &=& h \left( \frac{3}{4} k_1^2  \phi^2 - m^2_0 \right),
\label{2d-60}\\
    w'' - \frac{w'}{x} &=& w \left( \frac{3}{4} k_1^2  \phi^2 - m^2_3 \right),
\label{2d-70}
\end{eqnarray}
here we redefined $\phi /\phi_0 \rightarrow \phi$, $f /\phi_0  \rightarrow f$,
$v /\phi_0  \rightarrow v$, 
$h /\phi_0  \rightarrow h$, 
$w /\phi_0  \rightarrow w$, 
$\mu_1 /\phi_0 \rightarrow \mu_1$,
$M_{0,1} /\phi_0  \rightarrow M_{0,1}$, 
$m_{0,3} /\phi_0  \rightarrow m_{0,3}$, $\rho \phi_0  \rightarrow x$;
$\phi_0 = \phi(0)$.

\subsection{Numerical investigation}
\label{numerical1}

For the numerical calculations we choose the following parameters values 
\begin{equation}
	k_1 = \sqrt{\frac{4}{3}}, \quad 
	\phi (0) = 1, \quad 
	\lambda_1 = 0.1, \quad 
	v(0) = 0.5, \quad 
\label{2e-10}
\end{equation}
We apply the methods of step by step approximation for finding of numerical solutions (the details of similar calculations can be found in Ref. \cite{Dzhunushaliev:2003sq}). 
\par 
\textbf{Step 1}. On the first step we solve Eq. \eqref{2d-30} (having zero approximations $f_0(x) = \frac{0.2}{\cosh^2\left( \frac{x}{4} \right)}$), 
$\phi_0(x) = 1.3 - \frac{0.3}{\cosh^2\left( \frac{x}{4} \right)}$. The regular solution exists for a special value $M^*_{1,i}$ only. For $M_1 < M^*_{1,i}$ the function 
$v_i(x) \rightarrow +\infty$ and for $M_1 > M^*_{1,i}$ the function 
$v_i(x) \rightarrow -\infty$ (here the index $i$ is the approximation number). One can say that in this case we solve \emph{a non-linear eigenvalue problem}: $v_i^*(x)$ is the eigenstate and $M_{1,i}^*$ is the eigenvalue on this Step. 
\par 
\textbf{Step 2}. On the second step we solve Eq. \eqref{2d-40} using zero approximation $\phi_0(x)$ for the function $f(x)$ and the first approximation $v_1^*(x)$ for the function $v(x)$ from the Step 1. For $M_0 < M^*_{0,1}$ the function $f_1(x) \rightarrow +\infty$ and for $M_0 > M^*_{0,1}$ the function $f_1(x) \rightarrow -\infty$. Again we have 
\emph{a non-linear eigenvalue problem} for the function $f_1(x)$ and $M^*_{0,1}$.
\par 
\textbf{Step 3}. On the third step we solve Eq. \eqref{2d-50} using the first approximations $f^*_1(x)$ and $v^*_1(x)$ from the Steps 1,2. For $\mu_1 < \mu^*_{1,1}$ the function $\phi_1(x) \rightarrow +\infty$ and for $\mu_1 > \mu^*_{1,1}$ the function 
$\phi_1(x)$ oscillates and tends to zero. Again we have 
\emph{a non-linear eiqenvalue  problem} for the function $\phi_1(x)$ and $\mu^*_{0,1}$.
\par 
\textbf{Step 4}. On the fourth step we repeat the first three steps that to have the good convergent sequence $v_i(x), f_i(x), \phi_i^*(x)$. Practically we have made three approximations. 
\par 
\textbf{Step 5}. On the fifth step we solve Eq. \eqref{2d-60} using the first approximations $\phi^*_1(x)$ from the Step 3. This equation is exactly the Schr\"odinger equation for the function $h(x)$ with the potential $\phi^2(x)$ and eigenvalue $m_0^2$. 
\par 
\textbf{Step 6}. On the sixth step we solve Eq. \eqref{2d-70} using the first approximations $\phi^*_1(x)$ from the Step 3. For $m_3 < m^*_{3,1}$ the function 
$w_1(x) \rightarrow +\infty$ and for $m_3 > m^*_{3,1}$ the function 
$w_1(x) \rightarrow -\infty$. Again we have \emph{a non-linear eiqenvalue  problem} for the function $w_1(x)$ and $m^*_{3,1}$.
\par 
\textbf{Step 7}. On this step we repeat Steps 1-6 necessary number of times that to have the necessary accuracy of definition of the functions 
$v^*(x), f^*(x), \phi^*(x), h^*(x), w^*(x)$. 
\par
\begin{figure}[h]
  \begin{center}
  \fbox{
  \includegraphics[height=5cm,width=7cm]{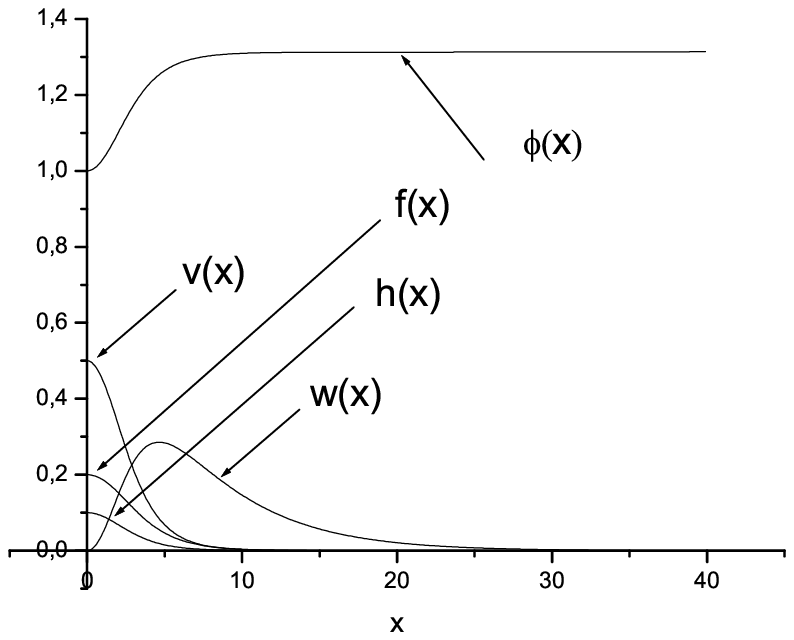}}
  \caption{The functions $f^*(x), v^*(x), h^*(x), w^*(x), \phi^*(x)$}
  \label{fig1}
  \end{center}
\end{figure}

\begin{figure}[h]
  \begin{center}
  \fbox{
  \includegraphics[height=5cm,width=7cm]{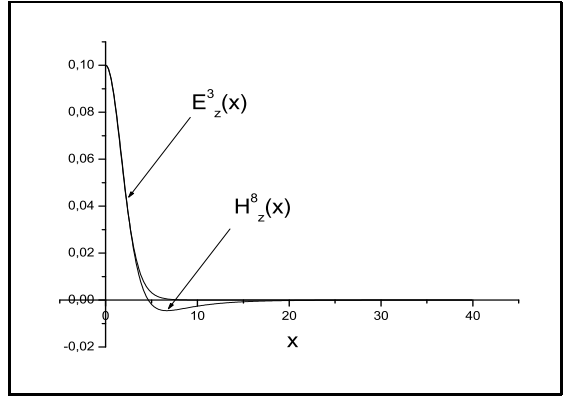}}
  \caption{The longitudinal chromoelectric $E^3_z(x)$ and chromomagnetic $H^8_z(x)$ fields.}
  \label{fig2}
  \end{center}
\end{figure}

\begin{figure}[h]
  \begin{center}
  \fbox{
  \includegraphics[height=5cm,width=7cm]{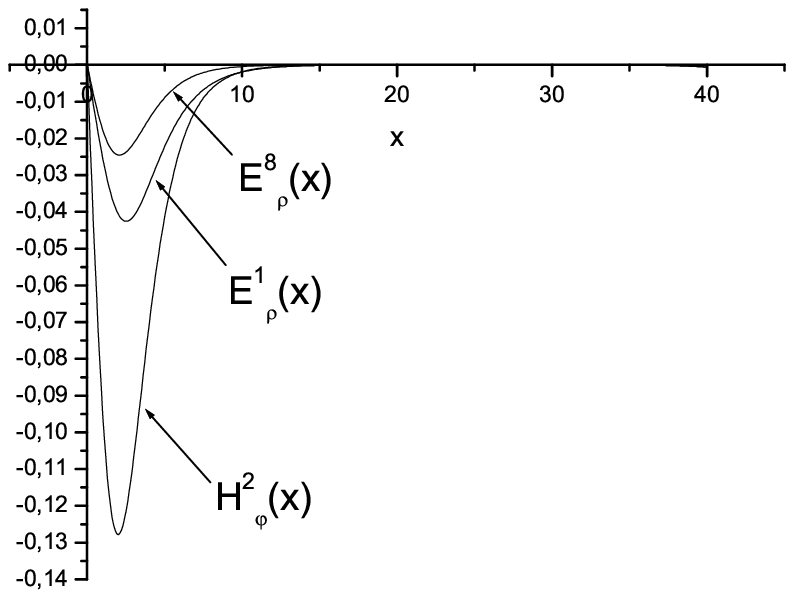}}
  \caption{The transversal chromoelectric $E^{1,8}_\rho(x)$ and chromomagnetic 			          $H^2_\varphi(x)$ fields.}
  \label{fig3}
  \end{center}
\end{figure}

 
After Step 7 we have the solution presented in Fig's. \ref{fig1} - \ref{fig3}. These numerical calculations give us the eigenvalues 
$M_0^* \approx 1.2325025678$, $M_1^* \approx 1.18060857823$, 
$m_0^* \approx 1.18356622792$, $m_3^* \approx 1.29306895$ 
and eigenstates $v^*(x), f^*(x), \phi^*(x), h^*(x), w^*(x)$. 


It is easy to see that the asymptotical behavior of the regular solution of 
equations \eqref{2d-30}-\eqref{2d-70} is
\begin{eqnarray}
    \phi(x) &\approx& \mu_1 + \phi_\infty \frac{\exp
    \left\{
    	-x\sqrt{2 \lambda_1 \mu_1^2}
    \right\}}{x},
\label{2e-20}\\
    v(x) &\approx& v_\infty \frac{\exp
    \left 
    	\{-x\sqrt{\mu_1^2 - M_0^2}
    \right\}}{x},
\label{2e-30}\\
    f(x) &\approx& f_\infty \frac{\exp
    \left 
    	\{-x\sqrt{\mu_1^2 - M_1^2}
    \right\}}{x},
\label{2e-40}\\
    h(x) &\approx& f_\infty \frac{\exp
    \left 
    	\{-x\sqrt{\mu_1^2 - m_0^2}
    \right\}}{x},
\label{2e-50}\\
    w(x) &\approx& f_\infty \frac{\exp
    \left 
    	\{-x\sqrt{\mu_1^2 - m_3^2}
    \right\}}{x}
\label{2e-60}
\end{eqnarray}
where $\phi_\infty, v_\infty, f_\infty, h_\infty, w_\infty$ are some constants. 
\par	
Let us note the following specialities of the solution: 
\begin{enumerate}
	\item 
	The flux of the chromoelectric field $E^3_z$ is nonzero 
\begin{equation}
	\Phi_E = 2 \pi \int \limits_0^\infty E^3_z \rho d \rho  = 
	2 \pi \int \limits_0^\infty f v \rho d \rho  \neq 0.
\label{2e-70}
\end{equation}
	\item
	The flux of the chromomagnetic field is zero 
\begin{equation}
	\Phi_H = 2 \pi \int \limits_0^\infty H^8_z \rho d \rho  = 
	2 \pi \int \limits_0^\infty \frac{d w}{d \rho} d \rho  = 
	2 \pi \left[ w(\infty) - w(0) \right] = 0.
\label{2e-80}
\end{equation}
	\item
	The longitudinal electric field 
\begin{equation}
	E^3_z = g A^1_t A^2_z
\label{2e-90}
\end{equation}
is essentially non-Abelian field.
	\item
	The longitudinal magnetic field 
\begin{equation}
	H^8_z = \frac{1}{\rho} \frac{d A^8_\phi}{d \rho}
\label{2e-100}
\end{equation}
is essentially Abelian field.
\item 
Comparing the obtained flux tube with the Nielsen - Olesen flux tube we see that the scalar field $\phi^a$ is an analog of the Ginzburg - Landau wave function for Cooper pairs and Eq. \eqref{2d-15} in QCD is the analog of the Ginzburg - Landau equation. 
\item
According to the previous remark the scalar field $\phi^a$ probably describes a monopole condensate. 
\end{enumerate}

\section{A defect with two infinitesimally closed electric and magnetic dipoles.}
\label{inft_close_dyon}

In the previous section we have obtained the color defect (infinite flux tube with the longitudinal electric field) where the color charges are on infinite distance from each other. In this section we want to investigate the case with infinitesimally closed color charges. 

\subsection{$SU(3) \rightarrow U(1) + coset$ decomposition}
\label{A}

Now we will consider the case when $A^{a,m}_\mu (a=1,2,3, m=4,5,6,7)$ degrees of freedom are quantized but $A^8_\mu$ remains in a classical phase. Again we start from the initial SU(3) Lagrangian $\mathcal L_{SU(3)}$ where we have to average over $A^{a,m}_\mu$. In order to do this we have to add assumptions in addition to the section \ref{basic_assumptions}:
\begin{enumerate}
\item 
On this step the gauge field components $A^{a}_\mu$ ($A^a_\mu \in SU(2)$) are in
a disordered phase ({\it i.e.} they form the gauge condensate), but have non-zero energy. In mathematical terms this means that
\begin{equation}
  \left\langle A^{a}_\mu (x) \right\rangle = 0.
\label{2f1-10}
\end{equation}
The reasonings similar to those which were resulted regarding section \ref{basic_assumptions} result us in expression 
\begin{equation}
  \left\langle A^a_\mu (x) A^b_\nu (y) \right\rangle = - \eta_{\mu \nu} 
  f^{apm} f^{bpn} \phi^m(x) \phi^n(y)
\label{2f1-20}
\end{equation}
here $\phi^m, m=4,5,6,7$ are real scalar fields. Thus we have replaced the SU(2) gauge fields by effective scalar fields, which will be the scalar fields in our effective 
$U(1)$+(two scalar field) system. 
\item
The 4-points Green's function for $A^a_\mu$ gauge potential can be decomposed by the following way 
\begin{eqnarray}
  \left\langle A^a_\mu(x) A^b_\nu(y) A^c_\alpha(z) A^d_\beta(u) \right\rangle &=&
  \lambda_2 \biggl[ \left\langle A^a_\mu(x) A^b_\nu(y) \right\rangle 
	  \left\langle A^c_\alpha(z) A^d_\beta(u) \right\rangle + 
  \biggl.,
\nonumber \\
	  &&
	  \frac{\mu_2^2}{4} \left(
	  	\delta^{ab} \eta_{\mu \nu} \left\langle A^c_\alpha(z) A^d_\beta(u) \right\rangle + 
	  	\delta^{cd} \eta_{\alpha \beta} \left\langle A^a_\mu(x) A^b_\nu(y) \right\rangle
	  \right) + 
\nonumber \\
	  &&
	  \frac{\mu_2^2}{4} \left(
	  	\delta^{ac} \eta_{\mu \alpha} \left\langle A^b_\nu(y) A^d_\beta(u) \right\rangle + 
	  	\delta^{bd} \eta_{\nu \beta} \left\langle A^a_\mu(x) A^c_\alpha(y) \right\rangle
	  \right) + 	
\nonumber \\
	  &&
	  \frac{\mu_2^2}{4} \left(
	  	\delta^{ad} \eta_{\mu \beta} \left\langle A^b_\nu(y) A^c_\alpha(u) \right\rangle + 
	  	\delta^{bc} \eta_{\nu \alpha} \left\langle A^a_\mu(x) A^d_\beta(y) \right\rangle
	  \right) +    ,
\nonumber \\
	  &&
	 \biggl.
	  \frac{\mu_2^4}{16} \left( 
	  \delta^{ab} \eta_{\mu \nu} \delta^{cd} \eta_{\alpha \beta} + 
	  \delta^{ac} \eta_{\mu \alpha} \delta^{bd} \eta_{\nu \beta} +
	  \delta^{ad} \eta_{\mu \beta} \delta^{bc} \eta_{\nu \alpha} 
	  \right)
	  \biggl] 
\label{2f1-30}
\end{eqnarray}
where $\lambda_2, \mu_2$ are constants. 
\item The correlations between the ordered ($A^8_\mu = b_\mu$) and disordered ($A^{a,m}_\mu$) phases have the following forms 
\begin{eqnarray}
  \left\langle A^{a,m}_\mu(x) A^{b,n}_\nu(x) b_\alpha (z) \right\rangle &=&
  k_1 b_\alpha (z) \left\langle A^{a,m}_\mu(x) A^{b,n}_\nu(x) \right\rangle ,
\label{2f1-32}\\
  \left\langle A^{a,m}_\mu(x) A^{b,n}_\nu(y) b_\alpha (z) b_\beta (u) \right\rangle &=&
  k_1^2 b_\alpha (z) b_\beta (u) \left\langle A^{a,m}_\mu(x) A^{b,n}_\nu(x) \right\rangle + 
  \delta^{mn} M_{1, \mu \nu} b_\alpha (z) b_\beta (u) ,
\label{2f1-34}
\end{eqnarray}
where $k_1$ are constant and $M_{1, \mu \nu}$ are matrices. 
\item 
We suppose that the Green's functions with odd numbers of $A^{a}_\mu$ are zero 
\begin{equation}
  \left\langle 
  	\overbrace{A^{a}_\mu (x) \ldots A^{b}_\nu (y)}^{\text{odd number}} 
  \right\rangle = 0 
\label{2f1-35}
\end{equation}
\item The correlations between $A^a_\mu$ and $A^m_\mu$ quantum degrees of freedom are  
\begin{eqnarray}
  \left\langle A^m_\mu(x) A^n_\nu(y) A^a_\alpha (z) A^b_\beta (u) \right\rangle &=&
  k_2^2 \left\langle A^m_\mu(x) A^n_\nu(x) \right\rangle 
  \left\langle A^a_\alpha (z) A^b_\beta (u) \right\rangle .
\label{2f1-36}
\end{eqnarray}
\end{enumerate}
Thus we have to calculate the following effective Lagrangian 
\begin{equation}
  \mathcal{L}_{eff} \approx 
  - \frac{1}{4}\left\langle \mathcal{F}^A_{\mu\nu}\mathcal{F}^{A\mu\nu}\right\rangle = 
	- \frac{1}{4} \left( 
		\left\langle \mathcal{F}^a_{\mu\nu}	\mathcal{F}^{a\mu\nu} \right\rangle + 
		\left\langle \mathcal{F}^m_{\mu\nu} \mathcal{F}^{m\mu\nu} \right\rangle + 
		\left\langle \mathcal{F}^8_{\mu\nu}	\mathcal{F}^{8\mu\nu} \right\rangle
	\right) 
\label{2f1-38}
\end{equation}
taking into account that $A^{a,m}_\mu$ are in disordered phase but $A^8_\mu$ remains in ordered phase.

\subsection{Calculation of
$\left\langle \mathcal{F}^a_{\mu\nu} \mathcal{F}^{a\mu\nu} \right\rangle$}

The calculations are similar with the section \ref{1a}. We begin by calculating the first term on the r.h.s. of equation \eqref{2f1-38}
\begin{equation}
\begin{split}
  \left\langle \mathcal{F}^a_{\mu\nu} \mathcal{F}^{a\mu\nu} \right\rangle = &
  \left\langle 
  	\left(
  		\partial_\mu A^a_\nu - \partial_\nu A^a_\mu 
  	\right)^2 
  \right\rangle +
  2 g f^{abc} \left\langle 
  	\left(
  		\partial_\mu A^a_\nu - \partial_\nu A^a_\mu 
  	\right) A^{b\mu} A^{c\nu}  
  \right\rangle +
  g^2 f^{abc} f^{ade} \left\langle 
  	A^b_\mu A^c_\nu A^{d \mu} A^{e \nu}
  \right\rangle + 
  \\
  &
  2 g f^{amn}
  \left\langle 
  	F^a_{\mu \nu} A^{m \mu} A^{n \nu}
  \right\rangle + 
  g^2 f^{amn} f^{apq} \left\langle 
  	A^m_\mu A^n_\nu A^{p \mu} A^{q \nu}
  \right\rangle
\label{2f2-10}
\end{split}
\end{equation}
The first term is 
\begin{equation}
  \left\langle 
  	\left(
  		\partial_\mu A^a_\nu - \partial_\nu A^a_\mu 
  	\right)^2 
  \right\rangle = 
  - \frac{3}{2} \partial_\mu \phi^m \partial^\mu \phi^m 
\label{2f2-15}
\end{equation}
The second term
$\left\langle 
  	\left(
  		\partial_\mu A^a_\nu - \partial_\nu A^a_\mu 
  	\right) A^{b\mu} A^{c\nu}  
  \right\rangle$  
on r.h.s. of equation \eqref{2c1-10} vanishes as it has odd number of $A^a_\mu$. 
The term $\left\langle F^a_{\mu \nu} A^{m \mu} A^{n \nu} \right\rangle$ is vanishes as $F^a_{\mu \nu}$ is the antisymmetrical tensor but 
$\left\langle A^{m \mu} A^{n \nu} \right\rangle$ is the symmetrical one (see Eq. \eqref{2f1-20}). 
\par
The two terms which are quartic in the coset gauge fields can be calculated using 
\eqref{2b-70} and \eqref{2f1-30}
\begin{eqnarray}
  g^2 f^{abc} f^{ade} \left\langle A^a_\mu A^b_\nu A^{d \mu} A^{e \nu} \right\rangle =
  \frac{9}{4} \lambda_2 g^2 \left(
  	\phi^m \phi^m - \mu_2^2
  \right)^2 ,
\label{2f2-20}\\
  f^{amn} f^{apq} \left\langle A^m_\mu A^n_\nu A^{p\mu} A^{q \nu} \right\rangle =
  \frac{9}{8} \lambda_1 g^2 \left(
  	\phi^a \phi^a - \mu_1^2
  \right)^2.
\label{2f2-30}
\end{eqnarray}
Up to this point the first term 
$\left\langle \mathcal{F}^a_{\mu\nu}	\mathcal{F}^{a\mu\nu} \right\rangle $ of the Lagrangian \eqref{2f1-38} is 
\begin{equation}
  \left\langle \mathcal{F}^a_{\mu\nu} \mathcal{F}^{a\mu\nu} \right\rangle =
  - \frac{3}{2} \partial_\mu \phi^m \partial^\mu \phi^m +
  \frac{9}{8} \lambda_1 g^2 \left(
  	\phi^a \phi^a - \mu_1^2
  \right)^2 +
  \frac{9}{4} \lambda_2 g^2 \left(
  	\phi^m \phi^m - \mu_2^2
  \right)^2.
\label{2f2-40}
\end{equation}

\subsection{Calculation of
$\left\langle \mathcal{F}^m_{\mu\nu} \mathcal{F}^{m\mu\nu} \right\rangle$}

The next term is 
\begin{equation}
\begin{split}
  \left\langle \mathcal{F}^m_{\mu\nu} \mathcal{F}^{m\mu\nu} \right\rangle = &
  \left\langle
		\left( \partial_\mu A^m_\nu - \partial_\nu A^m_\mu \right)
		\left( \partial^\mu A^{m \nu} - \partial^\nu A^{m \mu} \right)
  \right\rangle + 
  2 g f^{mna} \left\langle 
  	\left( \partial_\mu A^m_\nu - \partial_\nu A^m_\mu \right)
  	\left(
  		A^{a \nu} A^{n \mu} - A^{a \mu} A^{n \nu}
  	\right)
  \right\rangle + 
  \\
  &
    2 g f^{mn8} \left\langle 
  	\left( \partial_\mu A^m_\nu - \partial_\nu A^m_\mu \right) 
  	\left(
  		b^\nu A^{n \mu} - b^\mu A^{n \nu}
  	\right)
  \right\rangle + 
  \\
  &
  g^2 f^{mna} f^{mpb} \left\langle 
  	\left( A^a_\nu A^n_\mu - A^a_\mu A^n_\nu \right)
  	\left( A^{b\nu} A^{p \mu} - A^{b \mu} A^{p \nu} \right)
  \right\rangle + 
  \\
  &
  g^2 f^{mn8} f^{mp8} \left\langle 
  	\left( b_\nu A^n_\mu - b_\mu A^n_\nu \right)
  	\left( b^\nu A^{p \mu} - b^\mu A^{p \nu} \right)
  \right\rangle + 
  \\
  &
  g^2 f^{mna} f^{mp8} \left\langle 
  \left( A^a_\nu A^n_\mu - A^a_\mu A^n_\nu \right)
  \left( b^\nu A^{p \mu} - b^\mu A^{p \nu} \right)
  \right\rangle 
\end{split}
\label{2f3-10}
\end{equation}
After the calculations we have 
\begin{eqnarray}
	\left\langle
		\left( \partial_\mu A^m_\nu - \partial_\nu A^m_\mu \right)
		\left( \partial^\mu A^{m \nu} - \partial^\nu A^{m \mu} \right)
  \right\rangle &=& 
  -4 \left( \partial_\mu \phi^a \right) \left( \partial^\mu \phi^a \right) ,
\label{2f3-20}\\
	g^2 f^{mna} f^{mpb} 
	\left\langle 
		\left( A^a_\nu A^n_\mu - A^a_\mu A^n_\nu \right)
		\left( A^{b \nu} A^{p \mu} - A^{b \mu} A^{p \nu} \right)
	\right\rangle
	&=& 
  \frac{9}{4} g^2 k_2^2 \left( \phi^a \phi^a \right) 
  \left( \phi^m \phi^m \right)  ,
\label{2f3-30}\\
	g^2 f^{mn8} f^{mp8} 
	\left\langle 
		\left( b_\nu A^n_\mu - b_\mu A^n_\nu \right)
		\left( b^\nu A^{p \mu} - b^\mu A^{p \nu} \right)
	\right\rangle
	&=& 
  - 3 g^2 k_1^2 \left( \phi^a \phi^a \right) 
  b_\mu b^\mu + 2 \left( m^2 \right)^{\mu \nu} b_\mu b_\nu ,
\label{2f3-40}\\
	2 g f^{mna} 
	\left\langle 
		\left( \partial_\mu A^m_\nu - \partial_\nu A^m_\mu \right)
		\left( A^{a \nu} A^{n \mu} - A^{a \mu} A^{n \nu} \right)
	\right\rangle	&=& 0 ,
\label{2f3-50}\\
	g^2 f^{mna} f^{mp8} 
	\left\langle 
		\left( A^a_\nu A^n_\mu - A^a_\mu A^n_\nu \right)
		\left( b^\nu A^{p \mu} - b^\mu A^{p \nu} \right)
	\right\rangle
	&=& 0 ,
\label{2f3-60}\\
	2g f^{mn8} \left\langle 
		\left( \partial_\mu A^m_\nu - \partial_\nu A^m_\mu \right) 
		\left( b^\nu A^{n \mu} - b^\mu A^{n \nu} \right)
	\right\rangle &=& 0 .
\label{2f3-70}
\end{eqnarray}
Finally 
\begin{equation}
\begin{split}
  \left\langle
		\mathcal F^m_{\mu\nu} \mathcal F^{m\mu\nu}
  \right\rangle = &
  -4 \left( \partial_\mu \phi^a \right) \left( \partial^\mu \phi^a \right) 
  + \frac{9}{4} g^2 k_2^2 \left( \phi^a \phi^a \right) \left( \phi^m \phi^m \right)
  - 3 g^2 k_1^2 \left( \phi^a \phi^a \right) 
  b_\mu b^\mu + 
  2 \left( m^2 \right)^{\mu \nu} b_\mu b_\nu 
\end{split}
\label{2f3-80}
\end{equation}

\subsection{Calculation of
$\left\langle \mathcal{F}^8_{\mu\nu} \mathcal{F}^{8 \mu\nu} \right\rangle$}

We end by calculating the third term on the r.h.s. of equation \eqref{2f1-38}
\begin{equation}
  \left\langle \mathcal{F}^8_{\mu\nu} \mathcal{F}^{8 \mu\nu} \right\rangle =
  \left\langle h_{\mu\nu} h^{\mu\nu} \right\rangle +
  2 g \left\langle h_{\mu\nu} A^{m \mu} A^{n \nu} \right\rangle +
  g^2 f^{8mn} f^{8pq} 
  \left\langle 
  	A^m_\mu A^n_\nu A^{p \mu} A^{q \nu} 
  \right\rangle .
\label{2f4-10}
\end{equation}
The first term on the r.h.s. of \eqref{2c3-10} has the classical field $h_{\mu \nu}$ consequently 
\begin{equation}
  \left\langle h_{\mu\nu} h^{\mu\nu} \right\rangle = 
  h_{\mu\nu} h^{\mu\nu}.
\label{2f4-20}
\end{equation}
The second term in equation \eqref{2c3-10} vanishes as $h_{\mu \nu}$ is the antisymmetrical tensor but $\left\langle A^{m \mu} A^{n \nu} \right\rangle$ is the symmetrical one (see Eq. \eqref{2b-30}). The last term which is quartic in the coset gauge fields can be calculated using \eqref{2b-70}
\begin{equation}
  g^2 f^{8mn} f^{8pq} \left\langle A^m_\mu A^n_\nu A^{p\mu} A^{q \nu} \right\rangle =
  \frac{9}{8} \lambda_1 g^2 \left(
  	\phi^a \phi^a - \mu_1^2
  \right)^2.
\label{2f4-30}
\end{equation}
Thus 
\begin{equation}
  \left\langle \mathcal{F}^8_{\mu\nu} \mathcal{F}^{8\mu\nu} \right\rangle =
  h_{\mu\nu} h^{\mu\nu} +
  \frac{9}{8} \lambda_1 g^2 \left(
  	\phi^a \phi^a - \mu_1^2
  \right)^2.
\label{2f4-40}
\end{equation}

\subsection{An effective Lagrangian and field equations}

Thus after the quantization $A^{a,m}_\mu$ degrees of freedom we have 
\begin{equation}
\begin{split}
    -4 \left\langle \mathcal L_{eff} \right\rangle = & 
    h_{\mu \nu} h^{\mu \nu} 
    - \frac{3}{2} \left( \partial_\mu \phi^m \right) 
    \left( \partial^\mu \phi^m \right) 
    - 4 \left( \partial_\mu \phi^a \right) 
    \left( \partial^\mu \phi^a \right) +
    \frac{9}{4} \lambda_1 \left(
        \phi^a \phi^a - \mu_1^2 
    \right)^2 + 
    \frac{9}{4} g^2 \lambda_2 \left(
        \phi^m \phi^m - \mu_2^2 
    \right)^2 + 
    \\
    &
    \frac{9}{4} g^2 k_2^2 \left( \phi^a \phi^a \right)\left( \phi^m \phi^m \right) -
    3 g^2 k_1^2 \left( \phi^a \phi^a \right) b_\mu b^\mu + 
    2 \left( m^2 \right)^{\mu \nu} b_\mu b_\nu .
\end{split}
\label{4b-10}
\end{equation}
After the redefinition 
$\phi^m \rightarrow \frac{2}{\sqrt{3}} \phi^m,
\phi^a \rightarrow \frac{\phi^a}{\sqrt{2}},
\mu_2 \rightarrow \frac{2}{\sqrt{3}} \mu_2, 
\mu_1 \rightarrow \frac{\mu_1}{\sqrt{3}} \mu_1, 
4 g^2 \lambda_2 \rightarrow \lambda_2, 
\frac{9}{16} g^2 \lambda_1 \rightarrow \lambda_1$ 
we have 
\begin{equation}
\begin{split}
    \mathcal L_{eff} = & - \frac{1}{4} h_{\mu \nu} h^{\mu \nu} 
    + \frac{1}{2} \left( \partial_\mu \phi^a \right) 
    \left( \partial^\mu \phi^a \right) 
    + \frac{1}{2}  \left( \partial_\mu \phi^m \right) 
    \left( \partial^\mu \phi^m \right) -
    \frac{\lambda_1}{4} \left(
        \phi^a \phi^a - \mu_1^2 
    \right)^2 - 
    \frac{\lambda_2}{4} \left(
        \phi^m \phi^m - \mu_2^2 
    \right)^2 - 
    \\
    &
    \frac{3}{8} g^2 k_2^2 \left( \phi^a \phi^a \right)\left( \phi^m \phi^m \right) +
    \frac{3}{8} g^2 k_1^2 \left( \phi^a \phi^a \right) b_\mu b^\mu - 
    \frac{1}{2} \left( m^2 \right)^{\mu \nu} b_\mu b_\nu .
\end{split}
\label{4b-20}
\end{equation}
The field equations for the U(1)+(two scalar) system are
\begin{eqnarray}
    \partial^\mu \partial_\mu \phi^a &=& - \phi^a
    \left[
        \frac{3 k_2^2}{4} g^2 \phi^m \phi^m +
        \lambda_1 \left( \phi^a \phi^a - \mu_1^2 \right ) - 
        \frac{3}{4} g^2 k_1^2 b_\mu b^\mu
    \right] ,
\label{4b-30}\\
    \partial^\mu \partial_\mu \phi^m &=& - \phi^m
    \left[
        \frac{3 k_2^2}{4} g^2 \phi^a \phi^a +
        \lambda_2 \left( \phi^m \phi^m - \mu_2^2 \right )
    \right] ,
\label{4b-40}\\
    \partial_\nu h^{\mu \nu} &=&
    \frac{3}{4} g^2 k_1^2 \left( \phi^a \phi^a \right ) b^\mu - 
    \left( m^2 \right)^{\mu \nu} b_\nu .
\label{4b-50}
\end{eqnarray}

\subsection{Numerical solution}

\begin{figure}[h]
  \begin{center}
  \fbox{
  \includegraphics[height=5cm,width=7cm]{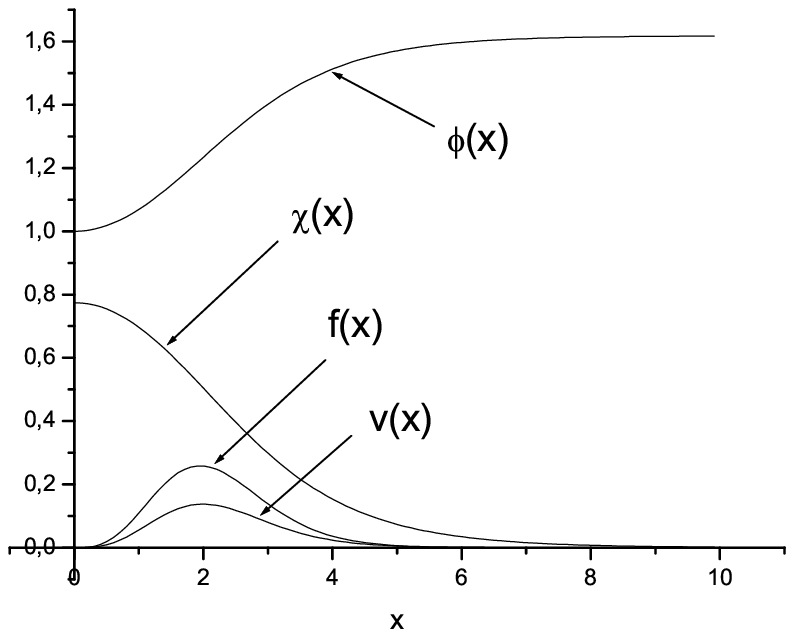}}
  \caption{The functions $f^*(x), v^*(x), \chi^*(x), \phi^*(x)$}
  \label{2f2-fig1}
  \end{center}
\end{figure}

We will use the following ans\"atz
\begin{equation}
\begin{split}
    &b_\mu = \sqrt{\frac{4}{3k_1^2}} \left\{ f(r, \theta), 0, 0, v(r, \theta) \right\}; 
    \quad 
    \phi^a = \frac{1}{g} \frac{2}{3k_2} \phi(r, \theta); 
    \quad 
    \phi^m = \frac{1}{g} \frac{1}{k_2\sqrt{3}} \chi(r, \theta) ;
    \\
    &
    \left( m^2 \right)^{\mu \nu} =
    \mathrm{diag} \left\{ m_0, 0, 0, \frac{m_3}{r^2 \sin^2\theta} \right\}
\label{4c-10}
\end{split}
\end{equation}
here $r, \theta , \varphi$ are spherical coordinates. The corresponding color (but Abelian) electric and magnetic fields are 
\begin{eqnarray}
    E^8_\theta &=& \frac{1}{r} \frac{\partial b_t}{\partial \theta} , \quad
    H^8_\theta = \frac{1}{r \sin \theta} \frac{\partial b_\phi}{\partial r},
\label{4c-20}\\
    E^8_r &=& - \frac{\partial b_t}{\partial r}, \quad
    H^8_r = \frac{1}{r^2 \sin \theta} \frac{\partial b_\phi}{\partial \theta}.
\label{4c-30}
\end{eqnarray}
Substituting \eqref{4c-10} into the U(1) - Higgs equations \eqref{4b-30}-\eqref{4b-50} gives us
\begin{eqnarray}
    \frac{1}{x^2} \frac{\partial}{\partial x}
    \left( x^2 \frac{\partial \phi}{\partial x} \right) +
    \frac{1}{x^2 \sin \theta} \frac{\partial}{\partial \theta}
    \left( \sin \theta \frac{\partial \phi}{\partial \theta}
    \right)
    &=& \phi
    \left[
      \chi^2 + \lambda_1 \left( \phi^2 - \mu^2_1 \right) -
      \left( f^2 - \frac{v^2}{x^2 \sin^2 \theta} \right)
    \right],
\label{4c-35}\\
    \frac{1}{x^2} \frac{\partial}{\partial x}
    \left( x^2 \frac{\partial \chi}{\partial x} \right) +
    \frac{1}{x^2 \sin \theta} \frac{\partial}{\partial \theta}
    \left( \sin \theta \frac{\partial \chi}{\partial \theta} \right)
    &=& \chi
    \left[
      \phi^2 + \lambda_2 \left( \chi^2 - \mu^2_2 \right)
    \right] ,
\label{4c-40}  \\
        \frac{1}{x^2} \frac{\partial}{\partial x}
    \left( x^2 \frac{\partial f}{\partial x} \right) +
    \frac{1}{x^2 \sin \theta} \frac{\partial}{\partial \theta}
    \left( \sin \theta \frac{\partial f}{\partial \theta} \right)
    &=& f \left[
        \left( \frac{k_1}{k_2} \right)^2 \phi^2 - m_0^2
    \right],
\label{4c-50}\\
    \frac{\partial^2 v}{\partial x^2} +
    \frac{\sin \theta}{x^2} \frac{\partial}{\partial \theta}
    \left( \frac{1}{\sin \theta}  \frac{\partial v}{\partial \theta} \right)
    &=& v \left[
        \left( \frac{k_1}{k_2} \right)^2 \phi^2 - m_3^2
    \right].
\label{4c-60}
\end{eqnarray}
here we redefined $\phi /\phi_0 \rightarrow \phi$, 
$\chi /\phi_0  \rightarrow \chi$,
$f /\phi_0  \rightarrow f$, 
$v /\phi_0  \rightarrow v$, 
$\frac{g k_2 \sqrt{3}}{2 \phi_0} \mu_1 \rightarrow \mu_1$, 
$\frac{g k_2}{\phi_0 \sqrt{12}} \mu_2 \rightarrow \mu_2$, 
$\frac{4}{3 g^2 k_2^2} \lambda_1 \rightarrow \lambda_1$, 
$\frac{12}{g^2 k_2^2 } \lambda_2 \rightarrow \lambda_2$, 
$r \phi_0  \rightarrow x$;
$\phi_0 = \phi(0)$. The equations \eqref{4c-35} - \eqref{4c-60} are partial differential equations and it is not quite clearly how they can be solved numerically to find eigenvalues $\mu_{1,2}$ and $m_{0,3}$. In order to avoid this problem we assume that 
the effect from the U(1) Abelian field $b_\mu$ is small and then 
\begin{eqnarray}
  \phi(x, \theta) &\approx & \phi(x), \quad 
  \chi(x, \theta) \approx \chi(x),
\label{2f2-60}  \\
	f(x, \theta) &=& f(x) \left( 
		\cos \theta - \frac{5}{3} \cos^3 \theta
	\right), 
\label{2f2-70}\\
	v(x, \theta) &=& v(x) \sin^2 \theta \cos \theta.
\label{2f2-80}
\end{eqnarray}
In addition we have to average Eq. \eqref{2f2-20} over the angle $\theta$ using the following average 
\begin{eqnarray}
  \frac{1}{\pi} \int \limits_0^\pi f^2(x, \theta) \sin \theta \; d \theta&=& 
  \frac{8}{63 \pi} f^2(x),
\label{2f2-90}  \\
  \frac{1}{\pi} \int \limits_0^\pi \frac{v^2(x, \theta)}{x^2} \sin \theta \; d \theta&=& 
  \frac{4}{15 \pi} \frac{v^2(x)}{x^2}.
\label{2f2-100}
\end{eqnarray}
Finally we have the following ordinary differential equations which can be numerically solved as an eigenvalue problem 
\begin{eqnarray}
  \phi '' + \frac{2}{x} \phi ' &=& 
  \phi \left[
  	\chi^2 + \lambda_1 \left( \phi^2 - \mu_1^2 \right) - 
  	\left( \frac{8}{63 \pi} f^2 - \frac{4}{15 \pi} \frac{v^2}{x^2} \right)
  \right],
\label{2f2-110}  \\
  \chi '' + \frac{2}{x} \chi ' &=& 
  \chi \left[
  	\phi^2 + \lambda_2 \left( \chi^2 - \mu_2^2 \right) 
  \right],
\label{2f2-120}\\
  f '' + \frac{2}{x} f ' - \frac{12}{x^2} f &=& 
  f \left[
  	\left( \frac{k_1}{k_2} \right)^2 \phi^2 - m_0^2
  \right],
\label{2f2-130}\\
  v '' - \frac{6}{x^2} v &=& 
  v \left[
  	\left( \frac{k_1}{k_2} \right)^2 \phi^2 - m_3^2
  \right].
\label{2f2-140}
\end{eqnarray}
For this choice we should have
\begin{eqnarray}
    \left.E_\theta(x, \theta)\right|_{\theta=0,\pi} &=& 0 , \quad
    \left.H_\theta(x, \theta)\right|_{\theta=0,\pi} = 0 ,
\label{2f2-150}\\
    \left.E_r(x, \theta)\right|_{x=0} &=& 0, \quad
    \left.H_r(x, \theta)\right|_{x=0} = 0 .
\label{2f2-155}
\end{eqnarray}
The series expansions near $x=0$ for the functions are 
\begin{eqnarray}
    \phi(x) &=& \phi_0 + \phi_3 \frac{x^2}{2} + \ldots ,
\label{2f2-160}\\
    \chi(x) &=& \chi_0 + \chi_3 \frac{x^2}{2} + \ldots ,
\label{2f2-170}\\
    f(x) &=& f_3 \frac{x^3}{6} + \ldots ,
\label{2f2-180}\\
    v(x) &=& v_3 \frac{x^3}{6} + \ldots
\label{2f2-190}
\end{eqnarray}
provide the constraints \eqref{2f2-150} \eqref{2f2-155}. We will search for a regular solution with the following boundary conditions:
\begin{eqnarray}
    \phi(0) &=& 1, \quad \phi(\infty) = \mu_1 ,
\label{2f2-195}\\
    \chi(0) &=& \chi_0, \quad \chi (\infty) = 0 ,
\label{2f2-200}\\
    f(0) &=& f(\infty) = 0,
\label{2f2-210}\\
    v(0) &=& v(\infty) = 0.
\label{2f2-220}
\end{eqnarray}
The numerical calculations are carried out as well as in the section \ref{numerical1} with $\frac{k_1^2}{k_2^2} = 6$. The result is presented in Fig \ref{2f2-fig1} and 
$\mu_1 = 1.617525 \ldots, \mu_2 = 1.4930404 \ldots, m_3 = 3.46539926281310275 \ldots$.

It is easy to see that the asymptotical behavior of the regular solution of
equations \eqref{2f2-110}-\eqref{2f2-140} is
\begin{eqnarray}
    \phi(x) &\approx& \mu_1 + \phi_\infty \frac{\exp \left\{ 
    	-x\sqrt{2 \lambda_1 \mu_1^2} 
    \right\}}{x},
\label{2f2-230}\\
    \chi(x) &\approx& \chi_\infty \frac{\exp 
    	\left\{ -x\sqrt{\mu_1^2 - \lambda_2 \mu_2^2}
    \right\}}{x},
\label{2f2-240}\\
    f(x) &\approx& f_\infty \frac{\exp 
    	\left\{ -x\sqrt{\left(\frac{k_1}{k_2}\right)^2 \mu_1^2 - m_0^2} 
    \right\}}{x},
\label{2f2-250}\\
    v(x) &\approx& v_\infty \exp 
    	\left\{-x\sqrt{\left(\frac{k_1}{k_2}\right)^2 \mu_1^2 - m_3^2} 
    \right\}
\label{2f2-260}
\end{eqnarray}
where $\phi_\infty, \chi_\infty, f_\infty, v_\infty$ are some constants. 
\par
\begin{figure}[h]
  \begin{minipage}[t]{.45\linewidth}
    \centering
    \fbox{
        \includegraphics[height=5cm,width=5cm]{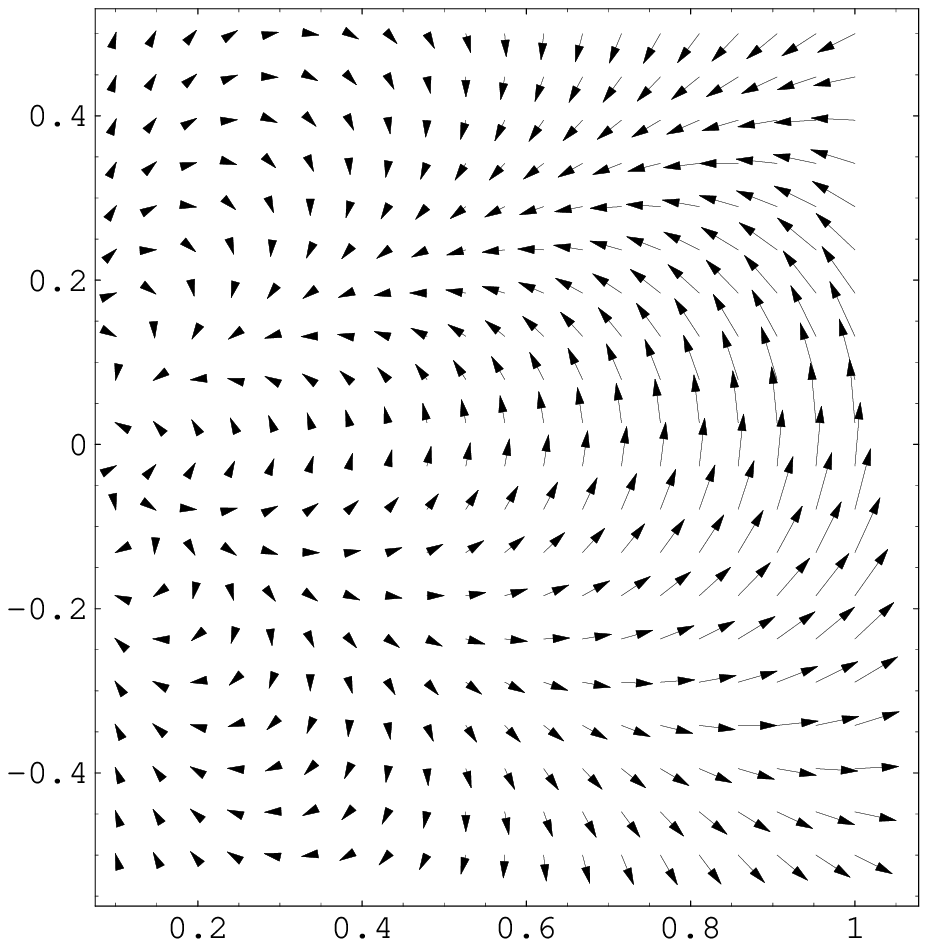}}
        \caption{The profile of the Abelian electric field $\vec E^8$.}
    \label{2f2-fig2}
  \end{minipage}\hfill
  \begin{minipage}[t]{.45\linewidth}
    \centering
    \fbox{
        \includegraphics[height=5cm,width=5cm]{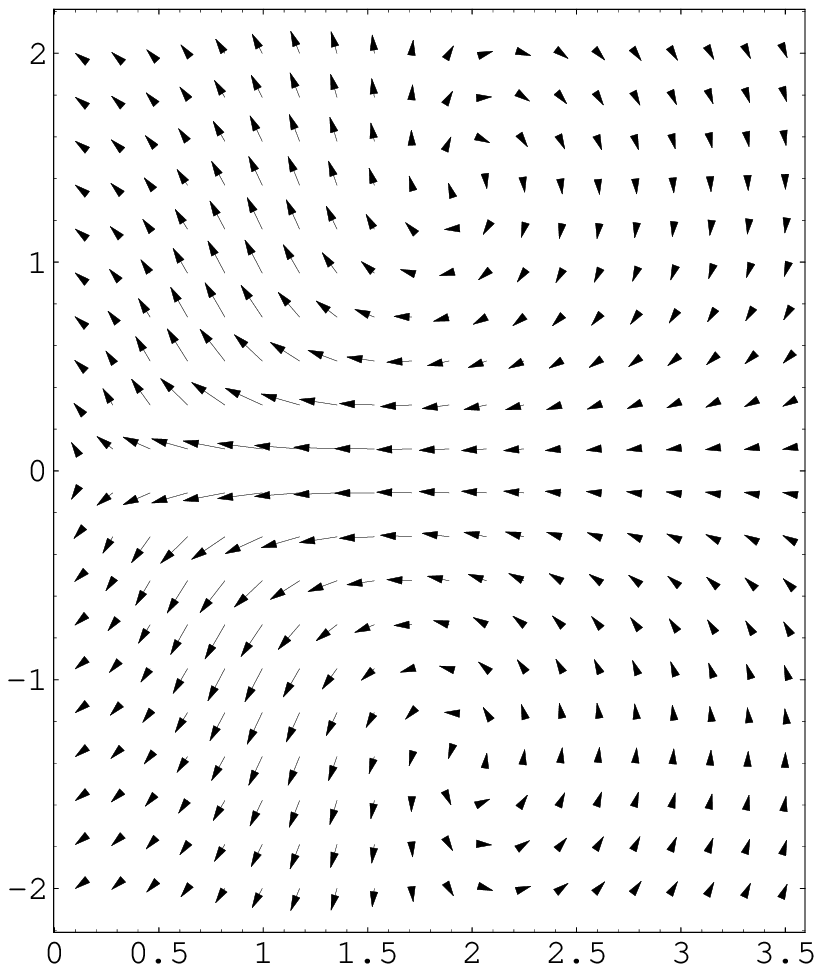}}
        \caption{The profile of the Abelian magnetic field $\vec H^8$.}
    \label{2f2-fig3}
  \end{minipage}
\end{figure}

 
In Fig. \eqref{2f2-fig2} and \eqref{2f2-fig3} the distribution of color electric and magnetic fields are presented.

\section{Color electric hedgehog}
\label{chedgehog}

In the previous section the defect in the gauge condensate is considered which represents two infinitesimally closed color electric dipoles and two color magnetic currents. 
In this section we want to investigate an isolated electric hedgehog in the gauge condensate. It is not pure electric charge as at the origin the radial electric field is zero. The origin of the electric hedgehog is the nonlinear interaction between ordered and disordered phases (nonlinear interaction between gauge potential components in the initial SU(3) Lagrangian). 
\par
\begin{figure}[h]
\begin{minipage}[t]{.45\linewidth}
\begin{center}
  \fbox{
  \includegraphics[height=5cm,width=7cm]{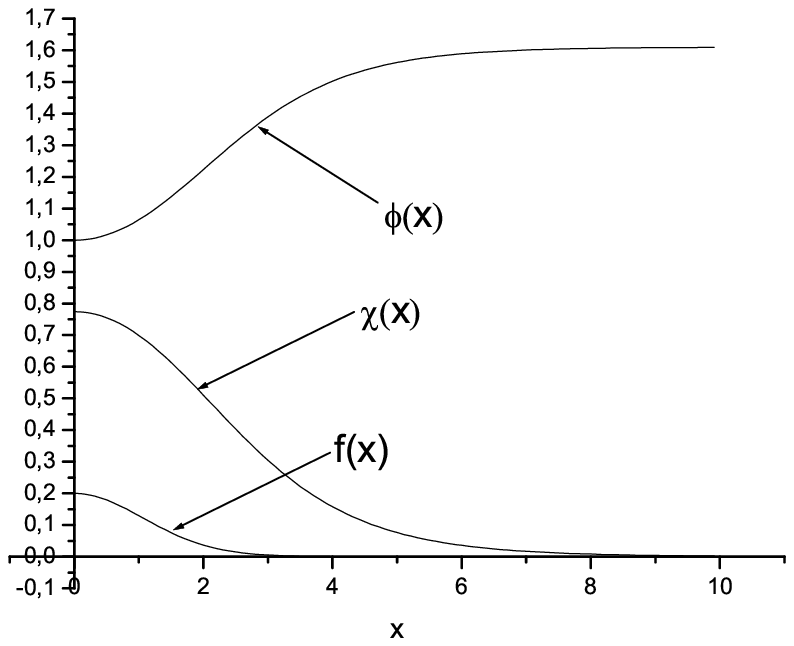}
  }
  \caption{The functions $\phi^*(x), \chi^*(x), f^*(x)$}
  \label{hdghhg_pot}
  \end{center}
\end{minipage}\hfill
\begin{minipage}[t]{.45\linewidth}
\begin{center}
  \fbox{
  \includegraphics[height=5cm,width=7cm]{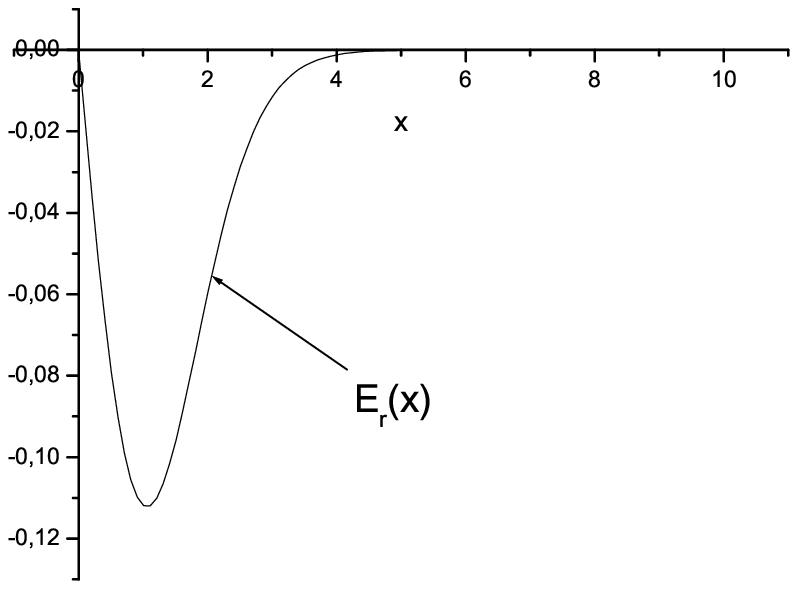}}
  \caption{The radial electric field $E_r$}
  \label{hdghhg_field}
  \end{center}
\end{minipage}
\end{figure}

The initial equations describing such defect are Eq's \eqref{4c-35}-\eqref{4c-60} with $v=0$ and we consider spherically symmetric case $f = f(r), \phi = \phi(r), \chi = \chi(r)$ 
\begin{eqnarray}
  \phi '' + \frac{2}{x} \phi ' &=& 
  \phi \left[
  	\chi^2 + \lambda_1 \left( \phi^2 - \mu_1^2 \right) - 
  	f^2 
  \right],
\label{3a2-10}  \\
  \chi '' + \frac{2}{x} \chi ' &=& 
  \chi \left[
  	\phi^2 + \lambda_2 \left( \chi^2 - \mu_2^2 \right) 
  \right],
\label{3a2-20}\\
  f '' + \frac{2}{x} f &=& 
  f \left[
  	\left( \frac{k_1}{k_2} \right)^2 \phi^2 - m_0^2
  \right].
\label{3a2-30}
\end{eqnarray}
The numerical calculations are the same as in the previous sections and the profiles of gauge components and radial electric field are presented in Fig's \ref{hdghhg_pot},  \ref{hdghhg_field}. In this case 
$\mu_1 = 1.6098972 \ldots, \mu_2 = 1.4854892 \ldots, m_0 = 2.9355398046322565 \ldots$

\section{Two color electric dipoles}

\begin{figure}[h]
\begin{center}
  \fbox{
  \includegraphics[height=5cm,width=7cm]{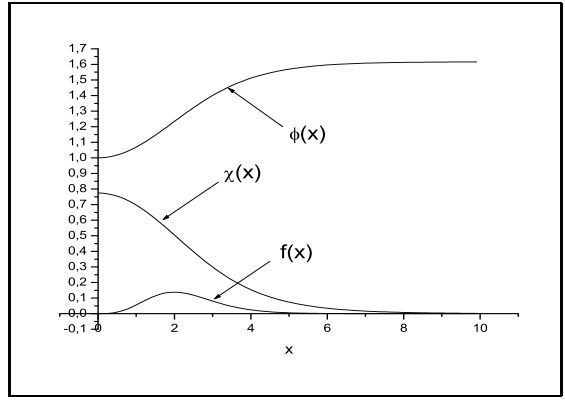}}
  \caption{The functions $\phi^*(x), \chi^*(x), f^*(x)$ for two color electric dipoles.}
  \label{el_dpl}
  \end{center}
\end{figure}

In this section we want to investigate Eq's \eqref{4c-35}-\eqref{4c-60} with $v=0$, \emph{i. e.} without magnetic field 
\begin{eqnarray}
    \frac{1}{x^2} \frac{\partial}{\partial x}
    \left( x^2 \frac{\partial \phi}{\partial x} \right) +
    \frac{1}{x^2 \sin \theta} \frac{\partial}{\partial \theta}
    \left( \sin \theta \frac{\partial \phi}{\partial \theta}
    \right)
    &=& \phi
    \left[
      \chi^2 + \lambda_1 \left( \phi^2 - \mu^2_1 \right) - f^2 
    \right],
\label{sec4-10}\\
    \frac{1}{x^2} \frac{\partial}{\partial x}
    \left( x^2 \frac{\partial \chi}{\partial x} \right) +
    \frac{1}{x^2 \sin \theta} \frac{\partial}{\partial \theta}
    \left( \sin \theta \frac{\partial \chi}{\partial \theta} \right)
    &=& \chi
    \left[
      \phi^2 + \lambda_2 \left( \chi^2 - \mu^2_2 \right)
    \right] ,
\label{sec4-20}  \\
        \frac{1}{x^2} \frac{\partial}{\partial x}
    \left( x^2 \frac{\partial f}{\partial x} \right) +
    \frac{1}{x^2 \sin \theta} \frac{\partial}{\partial \theta}
    \left( \sin \theta \frac{\partial f}{\partial \theta} \right)
    &=& f \left[
        \left( \frac{3k_1}{k_2} \right)^2 \phi^2 - m_0^2
    \right]
\label{sec4-30}
\end{eqnarray}
and 
\begin{equation}
		f(x, \theta) = f(x) \left( 
		\cos \theta - \frac{5}{3} \cos^3 \theta
	\right).
\label{sec4-40}
\end{equation}
Again we assume that the effect from the U(1) Abelian field $b_\mu$ is small and then 
\begin{equation}
  \phi(x, \theta) \approx \phi(x), \quad 
  \chi(x, \theta) \approx \chi(x).
\label{sec4-50}  
\end{equation}
Thus we will numerically investigate the following equations set
\begin{eqnarray}
  \phi '' + \frac{2}{x} \phi ' &=& 
  \phi \left[
  	\chi^2 + \lambda_1 \left( \phi^2 - \mu_1^2 \right) - 
  	\frac{8}{63 \pi} f^2 
  \right],
\label{sec4-60}  \\
  \chi '' + \frac{2}{x} \chi ' &=& 
  \chi \left[
  	\phi^2 + \lambda_2 \left( \chi^2 - \mu_2^2 \right) 
  \right],
\label{sec4-70}\\
  f '' + \frac{2}{x} f ' - \frac{12}{x^2} f &=& 
  f \left[
  	\left( \frac{k_1}{k_2} \right)^2 \phi^2 - m_0^2
  \right].
\label{sec4-80}
\end{eqnarray}
The result is presented in Fig. \ref{el_dpl} and 
$\mu_1 = 1.616718 \dots, \mu_2 = 1.4926212 \ldots,m_0 = 3.670495292 \ldots$ 
The field distribution of Abelian electric field $\vec E^8$ in this case is the same as in Fig. \ref{2f2-fig2} and similar to the field distribution between two electric dipoles lying on the line.

\section{Two color magnetic dipoles}

\begin{figure}[h]
\begin{center}
  \fbox{
  \includegraphics[height=5cm,width=7cm]{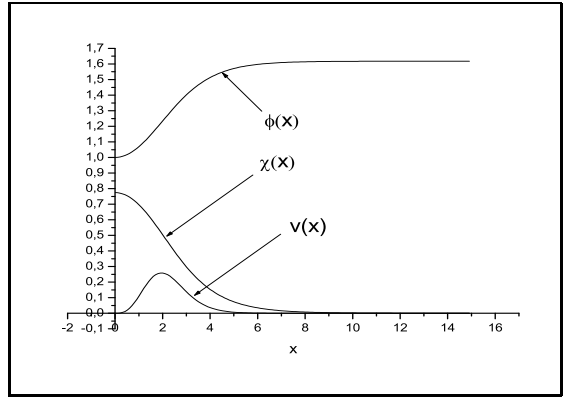}}
  \caption{The functions $\phi^*(x), \chi^*(x), v^*(x)$ for two color magnetic moments.}
  \label{mgn_mnts}
  \end{center}
\end{figure}

In this section we want to investigate Eq's \eqref{4c-35}-\eqref{4c-60} with $f=0$ 
\begin{eqnarray}
    \frac{1}{x^2} \frac{\partial}{\partial x}
    \left( x^2 \frac{\partial \phi}{\partial x} \right) +
    \frac{1}{x^2 \sin \theta} \frac{\partial}{\partial \theta}
    \left( \sin \theta \frac{\partial \phi}{\partial \theta}
    \right)
    &=& \phi
    \left[
      \chi^2 + \lambda_1 \left( \phi^2 - \mu^2_1 \right) + \frac{v^2}{x^2 \sin^2 \theta}        \right],
\label{sec5-10}\\
    \frac{1}{x^2} \frac{\partial}{\partial x}
    \left( x^2 \frac{\partial \chi}{\partial x} \right) +
    \frac{1}{x^2 \sin \theta} \frac{\partial}{\partial \theta}
    \left( \sin \theta \frac{\partial \chi}{\partial \theta} \right)
    &=& \chi
    \left[
      \phi^2 + \lambda_2 \left( \chi^2 - \mu^2_2 \right)
    \right] ,
\label{sec5-20}  \\
    \frac{\partial^2 v}{\partial x^2} +
    \frac{\sin \theta}{x^2} \frac{\partial}{\partial \theta}
    \left( \frac{1}{\sin \theta}  \frac{\partial v}{\partial \theta} \right)
    &=& v \left[
        \left( \frac{3k_1}{k_2} \right)^2 \phi^2 - m_3^2
    \right]
\label{sec5-30}
\end{eqnarray}
and 
\begin{equation}
	v(x, \theta) = v(x) \sin^2 \theta \cos \theta.
\label{sec5-40}  
\end{equation}
Doing the same assumptions about smallness of the U(1) Abelian field $b_\mu$ we have again  the approximations \eqref{sec4-50} and the following equations set
\begin{eqnarray}
  \phi '' + \frac{2}{x} \phi ' &=& 
  \phi \left[
  	\chi^2 + \lambda_1 \left( \phi^2 - \mu_1^2 \right) + \frac{4}{15 \pi} \frac{v^2}{x^2} 	  \right],
\label{sec5-50}  \\
  \chi '' + \frac{2}{x} \chi ' &=& 
  \chi \left[
  	\phi^2 + \lambda_2 \left( \chi^2 - \mu_2^2 \right) 
  \right],
\label{sec5-60}\\
  v '' - \frac{6}{x^2} v &=& 
  v \left[
  	\left( \frac{k_1}{k_2} \right)^2 \phi^2 - m_3^2
  \right].
\label{sec5-70}
\end{eqnarray}
The result is presented in Fig. \ref{mgn_mnts} and 
$\mu_1 = 1.6179587 \ldots, \mu_2 = 1.4931567 \ldots, m_1 = 3.465550906326 \ldots$. The field distribution of Abelian magnetic field $\vec H^8$ in this case is the same as in Fig. \ref{2f2-fig3} and similar to the field distribution between two magnetic dipoles.

\section{Parton-like defects in a gauge condensate}

In this paper we would like to include quark field. The initial effective Lagrangian is 
\begin{equation}
\begin{split}
    \mathcal L_{eff} = & - \frac{1}{4} h_{\mu \nu} h^{\mu \nu} 
    + \frac{1}{2} \left( \partial_\mu \phi^m \right) 
    \left( \partial^\mu \phi^m \right) 
    + \frac{1}{2}  \left( \partial_\mu \phi^a \right) 
    \left( \partial^\mu \phi^a \right) -
    \frac{\lambda_1}{4} \left(
        \phi^a \phi^a - \mu_1^2 
    \right)^2 - 
    \frac{\lambda_2}{4} \left(
        \phi^m \phi^m - \mu_2^2 
    \right)^2 - 
    \\
    &
    \frac{3}{8} g^2 k_2 \left( \phi^a \phi^a \right)\left( \phi^m \phi^m \right) +
    \frac{3}{8} g^2 k_1^2 \left( \phi^a \phi^a \right) b_\mu b^\mu - 
    \frac{1}{2} \left( m^2 \right)^{\mu \nu} b_\mu b_\nu + 
    \bar{\psi} \left[
    	\gamma^\mu \left(
    		\hat{p}_\mu + g A^8_\mu \lambda^8
    	\right) - m_q 
    \right] \psi 
\end{split}
\label{p3-10}
\end{equation}
where $\psi$ a quark field; $m_q$ is the quark mass; 
$\bar{\psi} \gamma^\mu A^B_\mu \lambda^B \psi = 
A^B_\mu \bar{\psi}_{i \alpha} \gamma^\mu_{\alpha \beta} \lambda^B_{ij} \psi_{j \beta}$. In Eq. \eqref{p3-10} we took into account that according to Section \ref{A}  
\begin{equation}
	\left\langle A^B_\mu \right\rangle = 
	\left\{
		\begin{array}{l}
			0,  		\;\;\; B=1,\cdots ,7 \\
			A^8_\mu,	B=8
		\end{array}
	\right.
\label{p3-20}
\end{equation}
Eq. \eqref{p3-20} means that $A^B_\mu, B=1,\cdots ,7$ are quantized degrees of freedom (disordered phase) but $A^8_\mu$ remains in classical state (ordered phase). The field equations for this system are
\begin{eqnarray}
  \partial^\mu \partial_\mu \phi^a &=& - \phi^a
  \left[
     \frac{3}{4} k_2^2 g^2 \phi^m \phi^m +
     \lambda_1 \left( \phi^a \phi^a - \mu_1^2 \right ) - 
     \frac{3}{4} g^2 k_1^2 b_\mu b^\mu
  \right] ,
\label{p3-30}\\
  \partial^\mu \partial_\mu \phi^m &=& - \phi^m
  \left[
     \frac{3}{4} k_2^2 g^2 \phi^a \phi^a +
     \lambda_2 \left( \phi^m \phi^m - \mu_2^2 \right )
  \right] ,
\label{p3-40}\\
  \partial_\nu h^{\mu \nu} &=&
  \frac{3}{4} g^2 k_1^2 \left( \phi^a \phi^a \right ) b^\mu - 
  \left( m^2 \right)^{\mu \nu} b_\nu + 
  g A^8_\mu \bar{\psi } \gamma^\mu \lambda^8 \psi ,
\label{p3-50}\\
	\left[
    \gamma^\mu \left(
    	\hat{p}_\mu + g A^8_\mu \lambda^8
    	\right) - m_q 
  \right] \psi &=& 0 .
\label{p3-60}
\end{eqnarray}
For the simplicity we choose $\psi_{1 \alpha} = \psi_{2 \alpha} = \psi_{3 \alpha}$. For this choice the term $\bar{\psi } \gamma^\mu \lambda^8 \psi$ in Eq. \eqref{p3-50} vanishes.
\par
\begin{figure}
\begin{center}
  \fbox{
  \includegraphics[height=5cm,width=7cm]{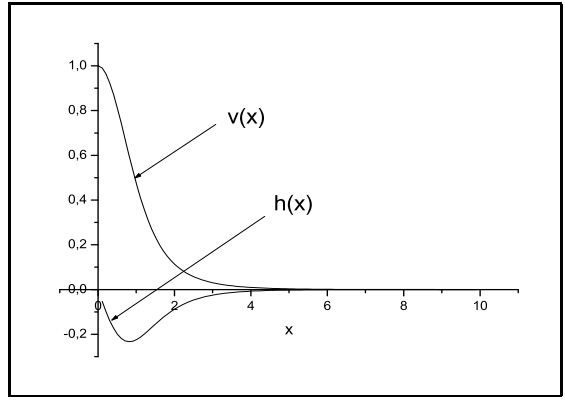}}
  \caption{The radial functions $v(r), h(r)$.}
  \label{pparton}
  \end{center}
\end{figure}

The solution of Eq's \eqref{p3-30}-\eqref{p3-60} we search in the form 
\begin{equation}
\begin{split}
    &b_\mu = \sqrt{\frac{4}{3k_1^2}} \left\{ f(r), 0, 0, 0 \right\}; 
    \quad 
    \phi^a = \frac{1}{g} \frac{2}{k_2} \phi(r); 
    \quad 
    \phi^m = \frac{1}{g} \frac{1}{k_2\sqrt{3}} \chi(r) ;
    \\
    &
    \left( m^2 \right)^{\mu \nu} =
    \mathrm{diag} \left\{ m_0, 0, 0, 0 \right\}
\label{p2-100}
\end{split}
\end{equation}
here $r, \theta , \varphi$ are spherical coordinates. The corresponding color (but Abelian) radial electric field is 
\begin{equation}
  E^8_r = - \frac{\partial b_t}{\partial r}.
\label{p2-110}
\end{equation}
The ans\"atz for the spinor field is in the standard form 
\begin{equation}
	\psi = 
	\left(
		\begin{array}{r}
			v(r) \Omega_{jlm} \\
			(-1)^{\frac{1+l-l'}{2}} h(r) \Omega_{jl'm} 
		\end{array}
	\right)
\label{p3-70}
\end{equation}
here $l = j+1/2$, $l' = 2j - 1$. Thus we have the following equations set
\begin{eqnarray}
  \phi '' + \frac{2}{x} \phi ' &=& 
  \phi \left[
  	\chi^2 + \lambda_1 \left( \phi^2 - \mu_1^2 \right) - 
  	f^2 
  \right],
\label{p3-80}  \\
  \chi '' + \frac{2}{x} \chi ' &=& 
  \chi \left[
  	\phi^2 + \lambda_2 \left( \chi^2 - \mu_2^2 \right) 
  \right],
\label{p3-90}\\
  f '' + \frac{2}{x} f &=& 
  f \left[
  	\left( \frac{k_1}{k_2} \right)^2 \phi^2 - m_0^2
  \right],
\label{p3-100}\\
  v' + \frac{1 + \varkappa}{x} v - 
  \left(
  	E + m - g f
  \right) h &=& 0 ,
\label{p3-110}\\
  h' + \frac{1 - \varkappa}{x} h + 
  \left(
  	E - m - g f
  \right) f &=& 0 
\label{p3-120}
\end{eqnarray}
here we have introduced the dimensionless coordinate $x = r\phi(0)$; redefined 
$v \rightarrow v/\phi_0, h \rightarrow h/\phi_0$ and 
\begin{equation}
	\varkappa = 
	\left\{
		\begin{array}{l}
			- \left( j+\frac{1}{2} \right) = - (l+1), 
			\;\;\; \text{if} \; 
			j=l+\frac{1}{2}\\
			+ \left( j+\frac{1}{2} \right) = l , 
			\quad \quad \quad \quad \text{if} \; 
			j=l - \frac{1}{2}
		\end{array}
	\right.
\label{p3-130}
\end{equation}
Eq's \eqref{p3-80}-\eqref{p3-100} have been solved in Section \ref{chedgehog}. The numerical calculations of Eq's \eqref{p3-110} \eqref{p3-120} (with $\varkappa = -1$) are presented in Fig. \ref{pparton}. 

\section{Discussion and conclusions}

In this work we have investigated color defects in a gauge condensate. The gauge condensate is defined as quantized non-perturbative degrees of freedom of the SU(3) gauge field (disordered phase). The color defects are residual degrees of freedom remaining in a classical state (ordered phase). Using some assumptions and approximations for 2 and 4-points Green's functions the SU(3) Lagrangian is reduced to an effective Lagrangian. The 2-point Green's function of quantum degrees of freedom is presented as a bilinear combination of scalar fields. The 4-point Green's function is presented as a bilinear combination of 2-points Greens functions. The effective Lagrangian describes a gauge theory with broken gauge symmetry interacting with scalar fields. The gauge field describe the ordered phase and scalar fields -- disordered phase. 
\par 
In this model we have a few undefined parameters which are connected with a non-linear interaction between initial gauge fields. Some of them are defined as eigenvalues at which regular solutions of corresponding field equations exist. The obtained solutions describe: 
infinite flux tube filled with longitudinal color electric and magnetic fields; a color electric hedgehog; a defect having two color electric dipoles lying on a line and color magnetic dipoles; two color electric dipoles; two color magnetic dipoles; parton-like defect which can be considered as a bag for the quark field. We assume that these defects are \emph{quantum excitations in a gauge condensate}. 
\par
Finally we would like to list the main results of the paper:
\begin{itemize}
\item 
An approximate method for non-perturbative quantization of QCD is proposed.
\item 
The proposed method is based on the decomposition of SU(3) degrees of freedom on ordered and disordered phases.
\item
The ordered phase is considered as classical gauge fields. This phase forms color defects inside of the disordered phase. For example, it can be: a flux tube filled with a longitudinal electric field, a hedgehog filled with a radial electric field and so on.
\item 
The disordered phase is quantum degrees of freedom which squeezes the ordered phase into color defects.
\item
Probably the disordered phase is formed by Abelian monopoles.
\item
The ordered phase is described by Yang-Mills equations with broken gauge symmetry.
\item
The equations for the disordered phase are an analog of Ginzburg - Landau equation for the wave function of Cooper pairs. 
\end{itemize}

\end{document}